\documentclass[a4paper,12pt]{article}
\usepackage{theorem}

\newtheorem{theorem}{Theorem}[section]

\newtheorem{lemma}[theorem]{Lemma}

\newcount\driver
\newcount\bozza

                    \font\ottorm=cmr8 scaled\magstep1
                  
\font\msytw=msbm10 scaled\magstep1                  
                 \font\indbf=cmbx10 scaled\magstep2

{\count255=\time\divide\count255 by 60 \xdef\hourmin{\number\count255}
        \multiply\count255 by-60\advance\count255 by\time
   \xdef\hourmin{\hourmin:\ifnum\count255<10 0\fi\the\count255}}

\let\a=\alpha \let\b=\beta    \let\g=\gamma     \let\d=\delta     \let\e=\varepsilon
  \let\h=\eta     \let\th=\vartheta      \let\l=\lambda
\let\m=\mu    \let\n=\nu      \let\x=\xi        \let\p=\pi        \let\r=\rho
\let\s=\sigma \let\t=\tau     \let\f=\varphi    \let\ph=\varphi   \let\c=\chi
\let\ps=\psi   \let\o=\omega     
 \let\D=\Delta       \let\L=\Lambda    
             
\let\O=\Omega 

\def\PP{{\cal P}}\def\EE{{\cal E}}\def\VV{{\cal V}}
\def\CC{{\cal C}}\def\HH{{\cal H}}\def\WW{{\cal W}} 
\def\KK{{\cal K}} 
\def\TT{{\cal T}}
\def\RR{{\cal R}}\def\LL{{\cal L}}
\def\DD{{\cal D}}\def\GG{{\cal G}}

\def\pp{{\bf p}}\def\qq{{\bf q}}\def\xx{{\bf x}}
\def\yy{{\bf y}}\def\kk{{\bf k}}\def\nn{{\bf n}}
\def\zz{{\bf z}}\def\uu{{\bf u}}\def\vv{{\bf v}}\def\ww{{\bf w}}
 \def\bP{{\bf P}}
\def\tt{{\bf t}}

\def\ss{{\underline \sigma}}       \def\oo{{\underline \omega}}
\def\ee{{\underline \varepsilon}}  
          
          \def\ux{{\underline\xx}}
\def\uk{{\underline \kk}}          \def\uq{{\underline\qq}}         
\def\up{{\underline\pp}}
           \def\uy{{\underline\yy}}
\def\uz{{\underline \zz}}          \def\uo{{\underline \o}}
\def\us{{\underline \s}}           \def\xxx{{\underline \xx}}

\def\u0{{\underline 0}}

\def\RRR{\hbox{\msytw R}}

        \def\EE{\hbox{\msytw E}}

\let\dpr=\partial

\let\bs=\backslash
\let\==\equiv

\let\io=\infty
\let\0=\noindent

\def\*{{\hfill\break\null\hfill\break}}

\def\ie{\hbox{\it i.e.\ }}

\def\tilde#1{{\widetilde #1}}

\def\lft{\left}
\def\rgt{\right}

\def\la{{\langle}}
\def\ra{{\rangle}}

\def\tende#1{\,\vtop{\ialign{##\crcr\rightarrowfill\crcr
             \noalign{\kern-1pt\nointerlineskip}
             \hskip3.pt${\scriptstyle #1}$\hskip3.pt\crcr}}\,}
\def\otto{\,{\kern-1.truept\leftarrow\kern-5.truept\to\kern-1.truept}\,}

\def\defi{{\buildrel \;def\; \over =}}

\def\wh#1{\widehat{#1}}
\def\hat#1{\wh{#1}}
\def\sqt[#1]#2{\root #1\of {#2}}

\def\ha{{\widehat \a}}\def\hb{{\widehat \b}}
\def\hr{{\widehat \r}}\def\hv{{\widehat v}}
\def\hf{{\widehat \f}}\def\hW{{\widehat W}}\def\hH{{\widehat H}}
\def\hB{{\widehat B}}
\def\hK{{\widehat K}} \def\hW{{\widehat W}}\def\hU{{\widehat U}}
\def\hp{{\widehat \ps}}

\def\hac{{\hat \c}}
\def\haf{{\hat f}}

\def\hJ{{\widehat J}}
\def\hg{{\widehat g}}       

\def\hQ{{\widehat Q}}
\def\hP{{\widehat P}}

\def\hA{{\widehat A}}

\def\hG{{\widehat G}}
\def\hS{{\widehat S}}
\def\hR{{\widehat R}}

\def\PP{{\cal P}}\def\EE{{\cal E}}\def\VV{{\cal V}}
\def\CC{{\cal C}}\def\HH{{\cal H}}\def\WW{{\cal W}}
\def\TT{{\cal T}}
\def\RR{{\cal R}}\def\LL{{\cal L}}
\def\DD{{\cal D}}\def\GG{{\cal G}}

\def\T#1{{#1_{\kern-3pt\lower7pt\hbox{$\widetilde{}$}}\kern3pt}}
\def\VVV#1{{\underline #1}_{\kern-3pt
\lower7pt\hbox{$\widetilde{}$}}\kern3pt\,}
\def\W#1{#1_{\kern-3pt\lower7.5pt\hbox{$\widetilde{}$}}\kern2pt\,}

\def\indica{\leaders \hbox to 0.5cm{\hss.\hss}\hfill}
\def\guida{\leaders\hbox to 1em{\hss.\hss}\hfill}
\mathchardef\oo= "0521

\def\pp{{\bf p}}\def\qq{{\bf q}}\def\xx{{\bf x}}
\def\yy{{\bf y}}\def\kk{{\bf k}}\def\nn{{\bf n}}
\def\zz{{\bf z}}\def\uu{{\bf u}}\def\vv{{\bf v}}
 \def\bP{{\bf P}}
\def\tt{{\bf t}} 
\def\ss{{\underline \sigma}}\def\oo{{\underline \omega}}

\def\xxx{{\underline\xx}}

\def\qed{\raise1pt\hbox{\vrule height5pt width5pt depth0pt}}

\def\indic{\hbox{\raise-2pt \hbox{\indbf 1}}}

\def\RRR{\hbox{\msytw R}}

\def\ins#1#2#3{\vbox to0pt{\kern-#2 \hbox{\kern#1 #3}\vss}\nointerlineskip}

\newdimen\xshift \newdimen\xwidth \newdimen\yshift
\newcount\griglia

\def\insertplot#1#2#3#4#5#6{%
\xwidth=#1pt \xshift=\hsize \advance\xshift by-\xwidth \divide\xshift by 2%
\begin{figure}[ht]
\vspace{#2pt}
\hspace{\xshift}
\begin{minipage}{#1pt}
#3
\ifnum\driver=1 \griglia=#6
\ifnum\griglia=1
\openout13=griglia.ps
\write13{gsave .2 setlinewidth}
\write13{0 10 #1 {dup 0 moveto #2 lineto } for}
\write13{0 10 #2 {dup 0 exch moveto #1 exch lineto } for}
\write13{stroke}
\write13{.5 setlinewidth}
\write13{0 50 #1 {dup 0 moveto #2 lineto } for}
\write13{0 50 #2 {dup 0 exch moveto #1 exch lineto } for}
\write13{stroke grestore}
\closeout13
\includegraphics{griglia.ps}
\fi
\includegraphics{#4.ps}\fi%
\ifnum\driver=2 \fi
\end{minipage}
\caption{#5}
\end{figure}
}

\newdimen\shift \shift=0.1truecm
\def\lb#1{%
\ifnum\bozza=1
\label{#1}\hbox{\hskip\shift$\scriptstyle#1$}
\else\label{#1}
\fi}

\def\be{\begin{equation}}
\def\ee{\end{equation}}
\def\bea{\begin{eqnarray}}\def\eea{\end{eqnarray}}
\def\bean{\begin{eqnarray*}}\def\eean{\end{eqnarray*}}
\def\bfr{\begin{flushright}}\def\efr{\end{flushright}}
\def\bc{\begin{center}}\def\ec{\end{center}}
\def\bal{\begin{align}}\def\eal{\end{align}}
\def\ba#1{\begin{array}{#1}} \def\ea{\end{array}}
\def\bd{\begin{description}}\def\ed{\end{description}}
\def\bv{\begin{verbatim}}\def\ev{\end{verbatim}}
\def\nn{\nonumber}
\def\Halmos{\hfill\vrule height10pt width4pt depth2pt \par\hbox to \hsize{}}
\def\pref#1{(\ref{#1})}

\begin{document}
\title{Renormalization Group and
Asymptotic Spin--Charge separation for Chiral Luttinger liquids}
\author{Pierluigi Falco \and  Vieri Mastropietro}
\date{}
\maketitle
\begin{center}
 Dipartimento di Matematica, Universit\`a di Roma ``Tor Vergata''
\\via della Ricerca Scientifica, I-00133, Roma
\end{center}
\begin{abstract}
The phenomenon of {\it Spin-Charge separation} in non-Fermi liquids  
is well understood only in certain {\it solvable} 
$d=1$ fermionic systems. 
In this paper we furnish 
the first example of {\it asymptotic} Spin-Charge separation in a $d=1$ {\it non solvable} model.
This goal is achieved using Renormalization Group approach combined with  
Ward-Identities and Schwinger-Dyson equations, corrected by  the presence
of a bandwidth cut-offs. Such methods, contrary to bosonization, 
could be in principle applied also to lattice or higher dimensional systems.
\end{abstract}

\section{Introduction and Main results}
In recent years the properties of non-Fermi liquids
have been extensively investigated, both from 
experimental and theoretical point of view. 
In particular, one of the most spectacular feature appearing in non-Fermi liquids 
is the phenomenon of {\it Spin-Charge (SC) separation}, which is
surely relevant for the physics of metals so anisotropic to 
be considered one dimensional, see for instance \cite{[S]} or \cite{[G]}. In addition, 
it is the key property in the Anderson theory
of high-$T_c$ superconductors (cuprates described by d=2 fermionic systems),
\cite{[A]}.

As it is well known,  SC separation is an 
highly non-perturbative phenomenon, and its occurrence in fermionic
models is quite hard to prove. Up to now it has been obtained only 
for the {\it spinning Luttinger model}
(or {\it Mattis model}), \cite{[Ma]}, describing 
two kind of fermions, with
spin $1/2$ and interacting through a short ranged potential. Its exact
solvability is due to the linear dispersion relation
(without any form of high energy cutoff) requiring a ``Dirac sea'' of fermions
with negative energy; such features are
quite unrealistic in a model aiming to
describe conduction electrons in a metal, but they allow to 
map the interacting fermionic system into a non-interacting
bosonic one, and to write the Hamiltonian as sum of two,  
decoupled Hamiltonians, respectively for spin and the charge 
degree of freedom. As a result, the two-point
Schwinger function, in the case of local interaction, 
factorizes into the product of two functions, 
with different {\it Fermi velocities},  $s_\r, s_\s$, and different
{\it critical indices} $\h_\r,\h_\s$, for the 
density ($\r$) and the spin ($\s$) respectively:
\be\lb{1a}
S_{\o}(x_0,x_1)={1\over (x_0 s_\s +i\o x_1)^{1/2+\h_\s}}
{1\over (x_0 s_\r +i\o x_1)^{1/2+\h_\r}}
\ee
Such a factorization appears also in the $n$-point Schwinger functions (see \cite{[M1]} for an explicit formula),
and it causes a phenomenology considerably different from the one of Fermi liquids \cite{[V]}.

For certain values of the parameters the spinning Luttinger
model reduces to the {\it Chiral Luttinger} model; in such a case
\pref{1a} still hold but $\h_\r=\h_\s=0$, that is in such a model 
only SC separation and no anomalous dimension is present.

The occurrence of SC separation
in more realistic {\it non solvable} models, like the Hubbard model,
has never been established, as a consequence of the fact that lattice or nonlinear bands prevents
the use of bosonization. It is important to understand
SC separation in the framework of Renormalization Group (RG), which is actually the only method
which can be in principle applied 
in full generality to the complex models appearing in condensed matter in any dimension. 
However even in $d=1$, in which RG methods have been extensively applied, 
- from the fundamental perturbative analysis in \cite{[S]}
to the non-perturbative and rigorous construction
of Luttinger liquids in \cite{[BGPS]}, \cite{[BM1]},\cite{[BM2]},\cite{[BM3]},\cite{[M2]}
- very few attention has been devoted to the application to SC separation effects
(with the exception of the recent paper \cite{[F]}, in which however
several approximations are introduced).

In this paper we will show that SC separation can be established in a {\it non exactly solvable} model
by using RG methods; the model we consider is the
{\it Chiral Luttinger liquid model} with a {\it bandwidth cut-off}, describing spinning fermions
interacting through a short range potential.
For physical applications, the presence of a {\it finite} momentum cut-off is essential
as a linear dispersion relation can be a reasonable approximation for a non-relativistic 
dispersion relation only for momenta close to the Fermi surface; its presence 
prevents however the possibility of an exact solution through bosonization.
This model have received a great deal of attention since
the edge excitations in the fractional Quantum Hall effect are believed to be a physical realization
of a Chiral Luttinger liquid \cite{[W]}.

\subsection{Basic definitions}

We express the Chiral Luttinger liquid model directly in terms  of
{\it Grassmann  variables}. Given the interval $[0,L]$, the inverse
temperature $\b$  and a large integer, $M$,  we introduce the lattice $\L_M$ 
made of the points $\xx=(x_0,x_1)=(n_0{\b\over M},n_1{L\over M})$,
for $n_0,n_1=0,1,\ldots M-1$.
We also consider the lattice  ${\cal D}={\cal D}_L \times {\cal D}_\b$
of points $\kk=(k_0,k_1)$, with
$k_0={2\pi\over \b} (n_0+{1\over 2})$, 
$k_1={2\pi\over L} (n_1+{1\over 2})$,
and $n_0,n_1=0,1,\ldots M-1$.
With each $\kk\in\DD$, we associate eight Grassmann variables, 
$\hp^{(\le N)\e}_{\kk,\o,\s}$, for $\e,\o,\s=\pm$; and we
consider the Grassman measure $P(d\ps^{(\le N)})$
that is defined in terms of the propagator 
$\la \ps^{-\e(\le N)}_{\xx,\o,\s}\ps^{\e'(\le N)}_{\yy,\o',\s'}\ra_0
=\e\d_{\e,\e'}\d_{\o,\o'}\d_{\s,\s'}
g_\o^{(\le N)}(\xx-\yy)$ for 
\be\lb{prop}
g_\o^{(\le N)}(\xx-\yy)
\defi
{1\over\b L}\sum_{\kk\in \DD} e^{i\kk(\xx-\yy)} {\hac_N(\kk)\over -i k_0+\o k_1}
\ee
where $\hac_N(\kk)$ is a smooth compact support function
$\hac_N(\kk)\defi \hac\lft(\g^{-N}|\kk|\rgt)$, where $\g>1$
and $\hac(t)$ is a 
$C^\io_0(\RRR_+)$ such that 
\be \hac(t)\defi
\lft\{\matrix{
1\hfill&\hfill{\rm if\ }0\le t\le 1\cr
0\hfill&\hfill{\rm if\ } t\ge \g.}\rgt.\ee
The {\it Generating Functional} of the
finite temperature Schwinger functions are obtained from 
the following {\it Grassman integral}
\bea\lb{gf1}\;
e^{\WW(\f,J)}\cr\cr
\defi 
\int\!P(d\ps^{(\le N)})\; 
\exp\Bigg\{\l V(\ps^{(\le N)})
+\sum_{\o,\s}
\int\!d\xx\;J_{\xx,\o,\s} 
\psi^{+(\le N)}_{\xx,\o,\s}\psi^{-(\le N)}_{\xx,\o,\s}\Bigg\}
\cr\cr\cdot 
\exp\Bigg\{\sum_{\o,\s}\int\!d\xx\;
\lft[\f^{+}_{\xx,\o,\s}\ps^{-(\le N)}_{\xx,\o,\s}+
\ps^{+(\le N)}_{\xx,\o,\s}\f^{-}_{\xx,\o,\s}\rgt]
\Bigg\}
\eea
where $\int d\xx\defi{\b L \over M^2}\sum_{x_0,x_1\in \L_M}$ and,
for $v(\xx)$ a smooth, rotation invariant, short range potential 
with $\hv(0)=1$, 
\be \lb{1.9}
V(\ps)=\sum_{\o,\s}
\int\!d\xx d\yy\; 
\ps^{+}_{\xx,\o,\s}\ps^{-}_{\xx,\o,\s}
v(\xx-\yy) \ps^{+}_{\yy,\o,-\s}\ps^{-}_{\yy,\o,-\s}\;.
\ee
$\{J_{\xx,\o}\}_{\xx,\o}$ are commuting variables, while
$\{\f^\s_{\xx,\o}\}_{\xx,\o,\s}$ are anticommuting.
The Schwinger functions are obtained 
by taking derivatives of $\WW(\f,J)$;
in particular the  two-point Schwinger function is defined 
\be
S_{N;\o,\s}(\xx-\yy)={\dpr^2 \WW_N\over 
\dpr\f^+_{\xx,\o,\s}\dpr\f^-_{\yy,\o,\s}}(0,0).
\ee

The lattice $\L_M$ is introduced just for technical reasons
in order to avoid an infinite number of Grassmann variables, but our results are trivially uniform
in $M$. 
The size $L$ and the inverse temperature $\b$
plays the role of infrared cut-offs; one is interested in the physical quantities
in the thermodynamic limit $L\to\io$ and at low temperatures, that is up to $\b=\io$.
We will prove the following result.
\begin{theorem}\lb{th3} 
There exists $\e_0>0$ ($N$ independent) such that,
for $|\l|\le\e_0$, the limit of the two-points Schwinger function for 
 $M,\b,L\to\io$ exists and has the form, for $\xx\not= {\bf 0}$
\be\lb{1x} 
S_{N;\o,\s}(\xx)={1\over (x_0 s +i\o x_1)^{1/2}(x_0 s^{-1} +i\o x_1)^{1/2}}
[1+\ R_N(\xx)]
\ee
with $R_N(\xx)$ bounded and such that 
\be
\lim_{|\xx|\to\io}R_N(\xx)=0
\qquad
{\rm and}\quad  s=1+{\l \over 2\pi}\;.
\ee
\end{theorem}
The above theorem provides  the first example
of SC separation in a {\it non solvable} model.
It is only {\it asymptotic}, that is up to terms which which are negligible for large distances.

The proof of \pref{1x} is based on Renormalization Group methods
combined with Ward Identities and Schwinger-Dyson equations, corrected 
by terms due to the presence of the momentum cut-offs which breaks the local symmetries.
Hopefully the methods presented here could be applied to 
prove spin-charge separation in the $d=1$ or even the $d=2$ Hubbard
model, despite such problems are of course much harder and pose
several  extra technical problems.

The rest of the paper is organized in the following way. In \S 2 and \S 3 we perform
a Renormalization Group analysis; in the integration of the ultraviolet scales
one has to improve the naive dimensional bounds taking advantage from the non-locality
of the interaction, while in the infrared scales dramatic cancellations
due to global phase symmetries are exploited.  In \S 4 we bound
the difference of the Schwinger functions with and without cut-offs, showing that it has a faster
power law decay. Finally in \S 5 we implement Ward Identities and Schwinger-Dyson
equations in the RG approach, obtaining an explicit expression of the Schwinger functions 
in the limit of removed cutoff.
\section{Renormalization Group analysis}\lb{s2}
We define the effective potential on scale $N$ 
\bea
\VV^{(N)}(\psi^{(\le N)},\f,J)
\defi
\l V(\ps^{(\le N)})
+\sum_{\o,\s}
\int\!d\xx\;J_{\xx,\o,\s} \psi^{+(\le N)}_{\xx,\o,\s}\psi^{-(\le
  N)}_{\xx,\o,\s}
\cr\cr
+\sum_{\o,\s}\int\!d\xx\;
\lft[\f^{+}_{\xx,\o,\s}\ps^{-(\le N)}_{\xx,\o,\s}+
\ps^{+(\le N)}_{\xx,\o,\s}\f^{-}_{\xx,\o,\s}\rgt]
\eea
Let  $\haf_h(\kk)\defi \hac\lft(\g^{-h}|\kk|\rgt) 
-\hac\lft(\g^{-(h-1)}|\kk|\rgt) $. The RG analysis is triggered by the
decomposition of 
$\hac_N(\kk)$ as $\sum_{h=-\io}^N \haf_h(\kk)$, and correspondingly, 
the decomposition of the propagator, \pref{prop}, as
\be
g^{(\le N)}_\o(\xx)=\sum_{h=-\io}^N g^{(h)}_\o(\xx)
\qquad
{\rm for}\quad g^{(h)}_\o(\xx)=
{1\over\b L}
\sum_{\kk\in \DD} e^{i\kk\xx} {\haf_h(\kk)\over -i k_0+\o k_1}\;.
\ee
Using standard techniques (see for instance \cite{[GM]}, appendix A2), 
for any positive integer $q$, there exists a 
constant $C_q$ such that, for any $h\le N$ 
\be\lb{17}
|g^{(h)}_\o(\xx)|\le C_q{\g^{h}\over 1+\big(\g^h|\xx|)^q}\;.
\ee
From the basic properties of Grassman integrals it also follows that
$\ps^{\e(\le N)}_{\xx,\o,\s}=\sum_{j=-\io}^N 
\ps^{\e(j)}_{\xx,\o,\s}$, where $\ps^{\e(j)}_{\xx,\o,\s}$ is randomly
independent from  $\ps^{\e(i)}_{\xx,\o,\s}$, for $i\neq j$; and has
covariance  $g^{(j)}_\o(\xx)$. We then define the {\it effective
  potential on scale $k$}, $\VV^{(k)}(\psi^{(\le k)},\f,J)$,  such that 
\bea\lb{2aa}\;
e^{\VV^{(k)}(\psi^{(\le k)},\f,J)}\defi
\int P(d\psi^{[k+1,N]}) e^{\VV^{(N)}(\psi^{[k+1,N]}
+\psi^{(\le k)},\f,J)}
\cr\cr
=e^{\sum_{n=1}^\io{1\over n!}\EE^T_{k+1,N} 
(\VV^{(N)};n)}
\eea
for $\ps^{\e[k, N]}_{\xx,\o,\s}=\sum_{j=k}^N 
\ps^{\e(j)}_{\xx,\o,\s}$ and $\ps^{\e(\le k)}_{\xx,\o,\s}=\sum_{j=-\io}^N 
\ps^{\e(j)}_{\xx,\o,\s}$;
$\EE_{k,N}^T$ is the {\it truncated expectation} with respect to
the propagator $g^{[k,N]}_\o(\xx)$:
$$
\EE^T_{k+1,N} (\VV^{(N)};n)\defi\EE^T_{k+1,N}
  [\underbrace{\VV^{(N)}|\cdots|\VV^{(N)}}_{n\ {\rm times}}]
$$
The effective potential is a polynomial of
the fields. For $\f=0$, (the case $\f\neq 0$ will be discussed in \S \ref{s4})  
we define the {\it kernels on scale $k$},
$W^{(n;2m)(k)}_{\o,\us}$, such that, for $\uz=\zz_1,\ldots, \zz_n$, 
$\ux=\xx_1,\ldots, \xx_m$,  $\uy=\yy_1,\ldots, \yy_m$ and
$\us=\s'_1,\ldots \s'_n, \s_1,\ldots \s_m$, we have

\bea\lb{ker}
\VV^{(k)}(\ps^{(\le k)},0,J)
=\sum_{n\ge 0\atop m\ge 0} 
\sum_{\us',\us\atop \o}\int\! d\uz d\ux d\uy\;
{W^{(n;2m)(k)}_{\o,\us}(\uz;\ux,\uy)\over n!(2m)!}
\cr\cr
\prod_{j=1}^n J_{\zz_i,\o,\s'_i}
\prod_{i=1}^{m}\ps^{+(\le k)}_{\xx_i,\o,\s_i}\ps^{-(\le k)}_{\yy_i,\o,\s_i}
\eea
As consequence of \pref{2aa}, the expression of the kernels in terms of the
truncated expectations is:
\bea\lb{3.29}
W^{(n;2m)(k)}_{\o,\us}
(\uz;\ux,\uy)
\cr\cr
=\lft.\prod_{i=1}^n
{\partial\over\partial J_{\zz_i,\o,\s'_i}}\rgt|_{J=0}
\lft.\prod_{i=1}^m
{\partial\over\partial \psi_{\xx_i,\o,\s_i}^{+(\le k)}}
{\partial\over\partial\psi_{\yy_i,\o,\s_i}^{-(\le k)}}\rgt|_{\psi^{(\le k)}=0}
\cr
\sum_{p=1}^\io {1\over p!}\EE_{k+1,N}^T(\VV^{(N)}(\psi^{(\le
  k)}+\psi^{[k+1,N]},J);p)
\eea
We introduce the following norm
\bea\lb{norm}
\|W^{(n;2m)(k)}_{\o,\us}\|_k
\cr\cr
\defi{1\over L\b}
\int\! d\ux d\uy d\ux' d\uy' d\uz\;
\lft|\underline\chi_k(\ux'-\ux) \underline\chi_k(\uy'-\uy)
W^{(n;2m)(k)}_{\o,\us}
(\uz;\ux,\uy)\rgt|
\eea
where $\underline\chi_k(\ux)=\prod_{j=1}^n \chi_k(\xx_j)$ and 
$\chi_k(\xx)$ is the Fourier transform of $\sum_{j\le k}\haf_j(\kk)$.

We give more details on the truncated expectation of monomials of the
fields; then, any polynomial can be  computed by multilinearity.
To shorten the notations we call
\be\label{lkf}
\psi_P=
\prod_{f\in P}\psi^-_{\xx(f),\o, \s(f)}\psi^+_{\yy(f),\o, \s(f)}
\ee
where $P$ is a set of labels.
Given the {\it clusters of points} $P_1, \ldots P_s$, 
the truncated expectation $\EE_{k+1,N}^T[\psi_{P_1}|\cdots|\psi_{P_s}]$
is given by the sum of the values (with the relative sign)
of all possible connected
Feynman graphs, obtained representing graphically 
the monomial $\psi_P$ as a set of oriented
half lines coming out from the clusters of points
and contracting them in all possible ways so that all the clusters are connected;
to each  line is associated a propagator $g^{[k+1,N]}_\o$.
\insertplot{210}{140}%
{\ins{95pt}{25pt}{$P_1$}
\ins{35pt}{85pt}{$P_2$}
\ins{135pt}{125pt}{$P_3$}
\ins{165pt}{65pt}{$P_4$}
}%
{n1} {\lb{n1}: An example of Feynman graph  corresponding to one
  possible  contribution to the
  truncated expectation of the clusters $P_1,\ldots,P_4$.
  The lines with the arrows are the propagator: not all of them are
  necessary to  connect the four clusters.}{0}
\* 
Then the  kernels $W^{(n;2m)(k)}_{\o,\us}$ can be written as sum over Feynman graphs as well,
and the presence of cutoffs make each of them finite.
Each connected Feynman
graph made of $p$ vertices is bounded by $C^p|\l|^p/p!$; anyway
their number is $O(p!^2)$ so that the sum of the graphs giving the
truncated expectations are bounded by $C^p|\l|^p p!$, 
from which convergence of the series expansion in $\l$ does not
follow. The combinatorial bound can be improved using the idea in 
\cite{[C]}:
the anticommutativity of fermions produces dramatic cancellations
among Feynman graphs, which are lost if the sum
of graphs is simply bounded by the sum of their absolute values.

In order to exploit such cancellations
it is then convenient to use a different representation of the
truncated expectations: here we follow the standard technique of
\cite{[GK]} and \cite{[FMRS]} (see also \cite{[L]} and, for a detailed
derivation, \cite{[GM]}).
\bea\lb{te2}
\EE^T_{k+1,N}[\psi_{P_1}|\ldots|\psi_{P_s}]
\cr\cr
=
\sum_{T}\prod_{l\in T} g^{[k+1,N]}_{\o}(\xx_l-\yy_l)
\int\!dP_{T}(\tt)\; 
 \det G^T_{k+1,N}(\tt)
\eea

where:
\bd
\item{1)} $T$ is a set of lines forming a {\it tree}
between the clusters of points $P_{1},\ldots,P_{s}$,
\ie $T$ is a set of lines which becomes a tree
if all the points in the same cluster are identified;
$n\defi \sum_{j=1}^s |P_j|$;
\item{2)}
$\tt=\{t_{i,i'}\in [0,1],
1\le i,i' \le s\}$ and $dP_{T}(\tt)$ is a probability measure with support
on a set of $\tt$ such that $t_{i,i'}=\uu_i\cdot\uu_{i'}$ for some family of
vectors $\uu_i\in \RRR^s$ of unit norm;
\item{3)}
$G^T_{k+1,N}(\tt)$ is a $(n-s+1)\times (n-s+1)$ matrix,
whose elements are given by
\be\lb{gd}
\left[ G^T_{k+1,N}(\tt) \right]_{(j,i),(j',i')}=t_{j,j'}
g^{[k+1,N]}_{\o}(\xx_{j,i}-\xx_{j',i'})
\ee
where $1\le j,j'\le s$ and $1\le i \le |P_{j}|$,
$1\le i'\le |P_{j'}|$, such that
the lines $l=\xx_{j,i}-\xx_{j',i'}$ do not belong to $T$.
\ed
\insertplot{210}{140}%
{\ins{95pt}{25pt}{$P_1$}
\ins{35pt}{85pt}{$P_2$}
\ins{135pt}{125pt}{$P_3$}
\ins{165pt}{65pt}{$P_4$}
}%
{n2}
{\lb{n2}:Graphical representation of one term in \pref{te2}. A tree
  graph connects  the four clusters. The determinant correspond to
  contract the remaining half lines each other in all possible ways}{0}
\*
The kernels $W^{(n;2m)(k)}_{\o,\us}$ can be written as a convergent
series in $\l$, as it is shown by the following lemma.
\begin{lemma}\lb{l1}
There exists $\e_{k,N}$ such that, for any $\l$ such that $|\l|\le \e_{k,N}$,
$W^{(n;2m)(k)}$ are analytic in $\l$.
\end{lemma}
{\bf\0Proof.}
We bound the determinant $G^T_{k+1,N}(\tt)$ in \pref{te2}
by using the
{\it Gram-Hadamard inequality}:
if $A_i$, $B_j$ are vectors in a Hilbert space with
scalar product $\la\cdot,\cdot\ra$, then
\be\lb{ghi}
|\det_{i,j} \la A_i,B_j\ra|\le \prod_i \sqrt{\la A_i, A_i\ra}\sqrt{\la
  B_i, B_i\ra}
\lb{7.27a}
\ee
Let $\HH=\RRR^s\otimes \HH_0$, where $\HH_0$ is the Hilbert space of
complex, squared summable functions, with scalar product
\be\label{7.27b}
\la F,G\ra=\sum_{i=1}^4 {1\over L\b}\sum_{\kk} \hat F^*_i(\kk) \hat G_i(\kk)
\ee
Since $G^T_{k+1,N}(\tt)$ in \pref{te2} can be written as
\bea\lb{7.45f}
G^{T}_{ij,i'j'}(\tt)=t_{i,i'}
g^{[h+1,N]}_{\o}(\xx_{ij}-\yy_{i'j'})
\cr\cr
=\la \uu_i\otimes A_{\xx_{ij},\o},
\uu_{i'}\otimes B_{\xx_{i'j'},\o}\ra
\eea
where $\uu_i\in \RRR^s$, $i=1,\ldots,s$, are the vectors such that
$t_{i,i'}=\uu_i\cdot\uu_{i'}$, and
\bea\label{[a[}
&&A_{\xx,\o} = {1\over L\b}\sum_{\kk}
e^{i\kk\xx}  { \sqrt{\hac_{k,N}(\kk)}
\over k_{0}^{2} + k^2 }\cr
&&B_{\xx,\o} = - {1\over L\b}\sum_{\kk} 
e^{i\kk\xx}   \sqrt{\hac_{k,N}(\kk)}
\, ( ik_{0} + \o k ) 
\eea
so that 
\be\label{dds}
\la A,A\ra^{1\over 2}\le C\g^{N-2k}\quad\quad  
\la B,B\ra^{1\over 2}\le C\g^{2N}\;,
\ee 
we get
\be\lb{17b}
|\det G^T_{k+1,N}(\tt)|\le 
C^{(\sum_{i=1}^s|P_i|/2-s+1) N} \g^{(\sum_{i=1}^s|P_i|/2-s+1) (N-k)}
\ee
The number of trees $T$ in \pref{te2} is bounded by $C^{\sum_i |P_i|} s!$,
for a suitable constant $C$; by using \pref{3.29} and \pref{te2}, 
bounding the determinants by \pref{17} and the integrations over coordinates by 
\be\label{fds}
\int\!d\xx\;
|g^{[h,N]}_\o(\xx)|\le C \g^{-h}
\quad\quad 
\int\!d\xx \;
|v(\xx)|\le C 
\ee
we get
\be\label{bion}
\|W^{(n,2m)(k)}_{\o,\us}\|_k\le 
\sum_{p=1}^\io |\l|^{p} C^p\g^{-p  3(N-k)}
\g^{m(p-3N)}\g^{-n k}\g^{3N+k}
\ee
and convergence follows for $\l$ small enough.
\qed

The above lemma says that the kernels $W^{(n,2m)(k)}_{\o,\us}$ are analytic in  $\l$ 
with an estimated radius of convergence which shrinks to zero when $|N-k|\to\io$; we will see
in the rest of this section how to improve the above bound to get convergence uniformly  
in $N-k$, by exploiting suitable cancellations in the series
expansion.

It is convenient to introduce the directional derivative
$$
\partial_\o={1\over 2}\lft(i{\partial\over\partial k_0}+\o
{\partial\over\partial k}\rgt)\;.
$$
We will skip, sometimes, the label
$\o$ in the kernels.
Calling $\hW^{(n;2m)(k)}_{\o,\us}(\up;\uk,\uq)$ the Fourier transform
of $W^{(n;2m)(k)}_{\o,\us}(\uz;\ux,\uy)$, we have the following lemma.
\begin{lemma}\lb{l2} 
For $|\l|$ small enough, 
\bea\lb{pcc}
\hW^{(0;4)(k)}_{\us}(0)=\l\d_{\s,-\s'}\;,
\quad
\hW^{(1;2)(k)}_{\us}(0)=\d_{\s,\s'}\;,
\cr\cr
\hW^{(0;2)(k)}_{\o,\us}(0)=\lft(\dpr_\o\hW^{(0;2)(k)}_{\o,\us}\rgt)(0)
=\lft(\dpr_{-\o}\hW^{(0;2)(k)}_{\o,\us}\rgt)(0)=0\;.
\eea
\end{lemma}
{\bf\0Proof.}
Because of lemma \ref{l1}, we can write the kernels as a convergent
power series in $\l$:
$\hW^{(n;2m)(k)}_{\o,\us}(\up;\uk,\uq)
=\sum_{p\ge 0}\l^p \hW^{(n;2m)(k)}_{p;\o,\us}(\up;\uk,\uq)$ .
For any  integer $p\ge 1$, we define $R_p\kk$ as the rotation of $\kk$
of an angle $\p\over 2p$:  
\be
\lft(
\begin{array}{c}
(R_p\kk)_0\cr\cr (R_p\kk)_1 
\end{array} \rgt)
=
\lft(
\begin{array}{cc}
\cos({\p\over 2p}) &-\sin({\p\over 2p})\cr\cr
\sin({\p\over 2p}) &\cos({\p\over 2p}) 
\end{array}\rgt)
\lft(
\begin{array}{c}
k_0\cr\cr k_1 
\end{array} \rgt)
\ee
so that, by the explicit expression of $\hg^{[k,N]}_\o$, and since
$\hv$ was defined invariant under rotations,
\be
\hg^{[k,N]}_\o(R_p\kk)
=e^{-i\o{\p\over 2p}} 
\hg^{[k,N]}_\o(\kk)\;,
\qquad
\hv(R_p\kk)=\hv(\kk)\;.
\ee
Since 
$\hW^{(0;4)(k)}_{p;\us}(\uk)$ 
is expressed by a sum over  connected Feynman graphs obtained contracting
$4p-4$ field (for such a kernel $p\ge 1$), we have
\be
\hW^{(0;4)(k)}_{p;\us}(R_p\uk)
=e^{-i\o\p(1-{1\over p})}\hW^{(0;4)(k)}_{p;\us}(\uk)\;,\ee
which implies $\hW^{(0;4)(k)}_{p;\us}(0)=0$
for any $p\ge 2$; while, for $p=1$,  $\hW^{(0;4)(k)}_{p;\us}(0)$ 
equals the coupling, $\l\d_{\s,-\s'}$.
In the same way $\hW^{(1;2)(k)}_{p;\us}(\uk)$ 
is sum over Feynman graphs obtained contracting
$4p$ fields (for $p\ge 0$); then 
\be
\hW^{(1;2)(k)}_{p;\us}(R_p\uk)=e^{-i\o\p}
\hW^{(1;2)(k)}_{p;\us}(\uk)
\ee  
and $\hW^{(1;2)(k)}_{p;\us}(0)=0$ for $p\ge 1$; while for $p=0$ 
$\hW^{(1;2)(k)}_{0;\us}(0)=\d_{\s,\s'}$. 
We also find
\bea\lb{rs2}
\hW^{(0;2)(k)}_{p;\s}(R_p\kk)
=
e^{-i\o\p(1-{1\over 2p})}
\hW^{(0;2)(k)}_{p;\s}(\kk)\;,
\cr\cr
\lft(\dpr_\o\hW^{(0;2)(k)}_{p;\o,\s}\rgt)(R_p\kk)
=
e^{-i\o\p}\lft(\dpr_\o\hW^{(0;2)(k)}_{p;\o,\s}\rgt)(\kk)\;,
\cr\cr
\lft(\dpr_{-\o}\hW^{(0;2)(k)}_{p;\o,\s}\rgt)(R_p\kk)
=
e^{-i\o\p(1-{1\over p})}\lft(\dpr_{-\o}\hW^{(0;2)(k)}_{p;\o,\s}
\rgt)(\kk)\;.
\eea
Since $p\ge 1$, and 
$\hW^{(0;2)(k)}_{1;\o,\s}(\kk)\= 0$ by explicit computation,
\pref{pcc} is proved. \qed
\*
We start now the multiscale integration. 
Using \pref{2aa}, we find 
\bea\lb{gf11}
e^{\WW_N(0,J)}
=
\int\!P(d\ps^{(\le N-1)})\int\!P(d\ps^{(N)}) 
e^{\VV^{(N)}(\psi^{(\le N)},0,J)}
\cr\cr
=
\int\!P(d\ps^{(\le N-1)})e^{\VV^{(N-1)}(\psi^{(\le N-1)},J)}
\eea
where $\VV^{(N-1)}(\ps^{(\le N-1)},0,J)$ has the same form of \pref{ker}. 
We introduce an $\LL$-operation
defined  on the kernels  in the following way
\bea\lb{if1}
\LL \hW^{(n;2m)(N-1)}_{\o,\us}(\uk)=0\quad {\rm if} \quad n+m> 2
\cr\cr
\LL \hW^{(n;2m)(N-1)}_{\o,\us}(\uk)=\hW^{(n;2m)(N-1)}_{\o,\us}(\uk)
\quad {\rm if}\quad 
n+m\le 2
\eea
Then we can write
\bea\lb{gf113}
e^{\WW_N(0,J)}=
\int\!P(d\ps^{(\le N-2)})
\cdot\cr\cr\cdot \int\!P(d\ps^{(N-1)})
e^{\LL\VV^{(N-1)}(\psi^{(\le N-1)},0,J)+\RR \VV^{(N-1)}(\psi^{(\le N-1)},0,J)}
\eea
and integrating we arrive to an expression similar to the r.h.s. of \pref{gf11}
with $N-1$ replaced by $N-2$; and so on for the integration of the $\ps^{(k+1)}$ field.
The above definition of $\LL$ remains the same until the  scale $k=0$.
For the fields on scales $k< 0$ we define:
\bea\lb{if11}
&& \LL\hW_{\us}^{(0;4)(k)}(\uk)
\defi\hW_{\us}^{(0;4)(k)}(0)\;,
\cr\cr
&&
 \LL\hW^{(1;2)(k)}_{\us}(\pp;\kk)
\defi
\hW^{(1;2)(k)}_{\us}(0;0)\;,
\cr\cr
&&\LL\hW^{(0;2)(k)}_{\o,\s}(\kk)
\defi
\hat W^{(0;2)(k)}_{\o,\s}(0)
+\kk\partial_\kk \hW^{(0;2)(k)}_{\o,\s}(0)
\eea
By lemma \ref{l2}, since 
$\kk\partial_\kk=
\sum_{\o'}D_{\o'}(\kk)\partial_{\o'}$,
we have that 
\bea\lb{ll}
&&\LL\hW_{\us}^{(0;4)(k)}(\kk,\pp,\qq)=\l\d_{\s,-\s'}\;,
\quad\LL\hW^{(1;2)(k)}_{\us}(\pp;\kk)=\d_{\s,\s'}\;,
\cr\cr
&&\LL\hW^{(0;2)(k)}_{\o,\s}(\kk)=0\;.
\eea
In performing the bounds, it is necessary to pass to the coordinate
representation; for $0\le k\le N$, we define $\l_{k;\o,\us}(\ux)$,
$\n_{k;\o,\s}(\ux)$ and 
$Z_{k;\o,\us}(\zz;\ux)$ such that 
\bea\lb{ef}
&&\LL\VV^{(k)}(\ps,0,J)=
\sum_{\o,\us}\int\!d\ux\; 
\l_{k;\o,\us}(\ux)
\psi^{+}_{\xx_1,\o,\s}\psi^{-}_{\xx_2,\o,\s}
\psi^{+}_{\xx_3,\o,\s'}\psi^{-}_{\xx_4,\o,\s'}
\cr\cr
&&+\sum_{\o,\s}
\int d\ux\;
\g^k\n_{k;\o,\s}(\ux)
\psi^{+}_{\xx_1,\o,\s}\psi^{-}_{\xx_2,\o,\s}
\cr\cr
&&+\sum_{\o,\us}\int\!d\zz d\ux\;
Z_{k;\o,\us}(\zz;\ux)J_{\zz,\o,\s}
\psi^{+}_{\xx_1,\o,\s'}\psi^{-}_{\xx_2,\o,\s'}
\eea
while for $k<0$
\bea\lb{loc} 
&&\LL\VV^{(k)}(\ps,0,J)=
\l\sum_{\us,\o}\int\!d\ux\; 
\d_{\s,-\s'}\d_3(\ux)
\psi^{+}_{\xx_1,\o,\s}\psi^{-}_{\xx_2,\o,\s}
\psi^{+}_{\xx_3,\o,\s'}\psi^{-}_{\xx_4,\o,\s'}
\cr\cr
&&+\sum_{\o,\us}\int\!d\zz d\ux\;
\d_{\s,\s'}\d_2(\zz,\ux)J_{\zz,\o,\s}
\psi^{+}_{\xx_1,\o,\s'}\psi^{-}_{\xx_2,\o,\s'}
\eea
where $\d_3(\ux)\defi \d(\xx_1-\xx_2)\d(\xx_2-\xx_3)\d(\xx_3-\xx_4)$
while $\d_2(\zz,\ux)\defi \d(\xx_1-\xx_2)\d(\xx_2-\zz)$.
To have a uniform notation we will also use the definitions,
for $k<0$, $\l_{k;\o,\us}(\ux)\defi \l\d_{\s,-\s'}\d_3(\ux)$ and 
$Z_{k;\o,\us}(\zz;\ux)\defi \d_{\s,\s'}\d_2(\zz,\ux)$.

It is well known, see for instance \cite{[BM1]}, that
$V^{(k)}(\psi^{(\le k)},0,J)$ can be represented as a 
sum over {\it Gallavotti-Nicol\`o trees} (in the following simply called
trees) defined in the following way.
\insertplot{300}{130}
{{\ins{30pt}{75pt}{$r$}
\ins{50pt}{75pt}{$v_0$}
\ins{130pt}{90pt}{$v$}%
\ins{35pt}{10pt}{$h$}
\ins{55pt}{10pt}{$h+1$}
\ins{135pt}{10pt}{$h_v$}%
\ins{235pt}{10pt}{$N$}
\ins{255pt}{10pt}{$N+1$}}%

}%
{n3}{\lb{n3}: A example of the  Gallavotti-Nicol\`o tree.}{0}
\*
The trees which can be constructed
by joining a point $r$, the {\it root}, with an ordered set of $n\ge 1$
points, the {\it endpoints} of the tree, 
so that $r$ is not a branching point. $n$ will be called the
{\it order} of the unlabeled tree and the branching points will be called
the {\it non trivial vertices}. We associate a label $h\le N-1$ with
the root, $r$ and we introduce
a family of vertical lines, labeled by an integer taking values
in $[h,N+1]$, and we represent any tree $\t\in\TT_{h,n}$ so that, if $v$ is an
endpoint or a non trivial vertex, it is contained in a vertical line with
index $h_v>h$, to be called the {\it scale} of $v$, while the root is on the
line with index $h$. 
The tree will intersect the vertical lines in set of points different
from the root and the endpoints;
these points will be called {\it trivial vertices}.
The set of the {\it
vertices} of $\t$ will be the union of the endpoints, the trivial vertices
and the non trivial vertices. Note that, if $v_1$ and $v_2$ are two vertices and $v_1<v_2$, then
$h_{v_1}<h_{v_2}$.
Moreover, there is only one vertex immediately following
the root, which will be denoted $v_0$ and can not be an endpoint;
its scale is $h+1$. There is the constraint, for the end-points of scale $h_v$,
that $h_v=h_{v'}+1$, if $v'$ is the first non trivial vertex immediately preceding $v$.
With each normal endpoint of scale $h_v$ we associate
$\LL\VV^{h_v-1}$ given by \pref{if1} if $h_v\ge 0$ or 
\pref{if11} if $h_v< 0$.

We introduce a {\it field label} $f$ to distinguish
the field variables appearing in the terms $\VV$
associated with the endpoints.
If $v$ is a vertex of the tree $\t$, $P_v$ is a set of
labels which distinguish the {\it external fields of $v$}, that is
the field variables of type $\psi$ which belong to one of the
endpoints following $v$ and are not yet contracted in the
vertex $v$. We will also call $n^\ps_v\defi |P_v|$ the number
of such fields $\psi$, and $n^J_v$ the number of the
field variables of type $J$ which belong to one of the
endpoints following $v$. 
Finally,  if $v$ is not an endpoint,  $\xx_v$ is the family of all
space-time points associated with one of the endpoints following
$v$.
It is easy to verify that
\be\lb{3.27}
V^{(k)}(\psi^{(\le k)},0,J) + \b L E_{k}=
\sum_{n\ge 1}\sum_{\t\in\TT_{k,n}}
V^{(k)}(\t)
\ee
where, if $v_0$ is the first vertex of $\t$ 
and $\t_1,\ldots,\t_s$ ($s=s_{v_0}$)
are the subtrees of $\t$ with root $v_0$,
$V^{(k)}(\t)$
is defined inductively by the relation, $k\le N-1$
\be\label{3.28}
V^{(k)}(\t)={(-1)^{s+1}\over s!} \EE^T_{k+1}[
\bar V^{(k+1)}(\t_1)|\cdots | \bar V^{(k+1)}(\t_{s})]
\ee
where $\bar V^{(k+1)}(\t)=\RR V^{(k+1)}(\t)$, for  $\RR=1-\LL$,
if the subtree $\t_i$ contains more then one endpoint;
if $\t_i$ contains only one endpoint $\bar
V^{(k+1)}(\t)$ is equal to one of the terms
in $\LL\VV^{h_v-1}$.

With these definitions, we can rewrite $\VV^{(k)}(\t,\psi^{(\le k)})$ as:
\bea\lb{3.29b}
&&\VV^{(k)}(\t)=\sum_{\bP\in\PP_\t}\VV^{(k)}(\t,\bP)
\cr\cr
&&\VV^{(k)}(\t,\bP)=\int d\xx_{v_0}
\psi^{(\le k)}_{P_{v_0}}
K_{\t,\bP}^{(k)}(\xx_{v_0})
\eea
where $K_{\t,\bP}^{(h+1)}(\xx_{v_0})$ is defined inductively 
by \pref{3.28}.

%
%
By lemma \ref{l1}  and 
calling $\e_{k}=\max_{\o,\us}\max_{k\le h\le N}\{
\|\l_{h;\o,\us}\|_k,\|\n_{h;\s}\|_k\}$
\be\lb{38}
\|K_{\t,\bP}^{(k)}\|_k\le (c \e_{k+1})^{n-n^J_{v_0}}
\g^{k(2-{|P_{v_0}|\over 2}-n^J_{v_0})}
\prod_{v\ {\rm not}\ {\rm e. p.}}\g^{-({|P_v|\over 2}-2+z_v+n^J_v)}
\ee
where, if $h_v>0$, $z_v\=0$. If $h_v\le 0$,  
$z_v=2$ if $|P_v|+2 n^J_v=2$; $z_v=1$ if $|P_v|+2 n^J_v=4$, and
$z_v=0$ otherwise.
The proof of \pref{38} is an immediate consequence of the analysis
in \S 3 of \cite{[BM1]}, based on \pref{te2} and teh Gram-Hadamard inequality.
The following lemma is an immediate consequence of the above bound.
\begin{lemma}\lb{l3}
There exist $C>1$ and $\e>0$ such that, for $\e_{k+1}\le \e$
and $\max_{h\ge k+1}\|Z_{k,\o,\s}\|_k<2$, 
\bea\lb{pc}
\|W^{(n;2m)(k)}_{\us,\uo}\|_k\le C^{n+m-1}\e^{(m-1\wedge 0)}
\g^{k(2-n-m)}.
\eea
for $(m\wedge 0)\defi \max\{m, 0\}$.
\end{lemma}
{\bf\0Proof.}
For $h_v>0$ the definition of $\RR$ imposes the constraint that
there are no $v$ such that $(|P_v|,n^J_v)=(4,0),(2,0),(2,1)$; this implies
that, for any $v$, 
\be\lb{rrr1}
d_v\defi{|P_v|\over 2}-2+z_v+n^J_v>0
\ee
In order to sum over $\t$ and $\bP$ (for more details, see again \cite{[BM1]})
we note that
the number of unlabeled trees is $\le
4^n$; fixed an unlabeled tree, the number of terms in the sum over the
various labels of the tree is bounded by $C^n$, except the sums over the scale
labels and the sets $\bP$. Let $V(\t)$ the nontrivial vertices of $\t$.
In order to bound the sums over the scale labels and $\bP$ 
we use the inequality
\bea
\prod_{v\ {\rm not}\ {\rm e. p.}}\g^{-({|P_v|\over 2}-2+z_v+n^J_v)}
=\prod_{v\in V(\t)}\g^{-(h_v-h_{v'}) d_v}
\cr\cr
\le 
\lft(\prod_{v\in V(\t)} 
\g^{-{1\over 40}(h_{v}-h_{v'})}\rgt)
\prod_{v\in V(\t)}\g^{-{|P_v|\over 40}}
\eea
and the first factor in the r.h.s. allow
to bound the sums over the scale labels by $C^n$,
while the the sum over $\bP$ can be bounded by using the following combinatorial
inequality. Let $\{p_v, v\in \t\}$ a set of integers such that
$p_v\le \sum_{i=1}^{s_v} p_{v_i}$ for all $v\in\t$ which are not endpoints;
then 
\be\label{gjj}
\sum_{\bP}\prod_{v\in V(\t)}
\g^{-{|P_v|\over 40}}\le \prod_{v\in V(\t)} \sum_{p_v}
\g^{-{p_v\over 40}} B\lft(\sum_{i=1}^{s_v}p_v,p_v\rgt) \le C^n
\ee
where $B(n,m)$ is the binomial coefficient. \qed
\section{Power counting improvement}

The bound \pref{pc} is of course not sufficient to prove
the boundedness of the kernels $W^{(n;2m)(k)}_{\us,\uo}$, as 
we need to prove that $\bar\e_k$ is small uniformly in $k$. On the other
hand $\vec v_h=(\l_h,\g^h \n_h,Z_h^{(2)})$ verify the equation, for $h\ge 0$ 
\be
\vec v_{h-1}=\vec v_h+\vec\b_h(\vec v_h,..,\vec v_N)
\ee
where $\vec\b_h$ is expressed by a sum of trees such that the first non trivial vertex
has scale $h+1$ (from the property $\LL\RR=0$), and $\vec v_N=(\l\d_{-\s',\s},0,\d_{\s,\s'})$.
Iterating the above equation one finds
\bea\lb{pc100}
\l_{h;\o,\us}(\ux)=W^{(0;4)(h)}_{\o,\us}(\ux)
\cr\cr
\g^h \n_{h;\o,\s}(\ux)=W^{(0;2)(h)}_{\o,\s}(\ux)\quad
Z_{h;\o,\us}(\zz;\ux)=W^{(1;2)(h)}_{\o,\us}(\zz;\ux)
\eea
and there is no reason a priori for which $\vec v_h$ should remain close
to $\vec v_N$; this property will be established by a careful analysis
implying an improvement of the previous bounds.
We will prove in fact the following theorem.

\begin{theorem}\lb{t3.2}
For $|\l|$ small enough, there exist a constant $C_1>1$ 
such that, for $0\le h\le N$ 
\bea\lb{hb}
\|W^{(0;2)(h)}_{\s}\|_h
\le C_1|\l|\g^{-h}\;,
\qquad
\|W^{(1;2)(h)}_{\s',\s}-\d_2\d_{\s,\s'}\|_h
\le C_1|\l|\g^{-h}\;,
\cr
\|W^{(0;4)(h)}_{\s,\s'}-v\l\d_2\d_{\s,-\s'}\|_h
\le C_1|\l|\g^{-h}\;;
\eea
where (with slight abuse of notation) $v\d_2\=\d(\xx-\yy)v(\xx-\uu)\d(\uu-\vv)$.
\end{theorem} 
\*
An immediate consequence of the above theorem, together
with \pref{if11}, \pref{ll},\pref{pc}, \pref{pc100} is the boundedness of the kernels
$W^{(n;2m)(k)}_{\o,\us}$ for $|\l|$ small enough (and since, for
$h\ge 0$, $\g^{-h}\le 1$)
\be
\lb{pc1}
 \|W^{(n;2m)(k)}_{\us,\uo}\|_k\le C^{n+m-1}|C_1 \l|^{(m-1\wedge 0)}
\g^{k(2-n-m)}  
\ee

{\bf\0Proof.}
The proof is by induction: we assume that \pref{hb} holds for
$h:k+1\le h\le N$; hence the hypothesis of lemma \ref{l3} are
satisfied and we can use \pref{pc} to prove \pref{hb} for $h=k$.

To shorten the notation, in this proof we call $\h\defi\ps^{\le k}$.
By definition of the effective interaction, $\VV^{(k)}$, 
we have 
\bea
W^{(n;2m)(k)}_{\uo,\us}(\uz;\ux,\uy)\lb{genb}
\cr\cr
={\partial^{n+2m}\VV^{(k)}\over
\dpr J_{\zz_1,\s_1}\cdots\dpr J_{\zz_n,\s_n}
 \dpr\h^+_{\xx_1,\o_1}\dpr\h^-_{\yy_1,\o_1}
 \dpr\h^+_{\xx_m,\o_m}\dpr\h^-_{\yy_m,\o_m}}(0,0,0)
\eea
%
%
By the explicit expression of the function $\VV^{(N)}$ we obtain: 
\bea
{\dpr \VV^{(k)}\over \dpr \h^+_{\xx,\o,\s}}(\h,J,0)
=J_{\xx,\o,\s}{\dpr \VV^{(k)}\over\dpr \f^+_{\xx,\o,\s}}(\h,J,0)
\lb{mas1} \phantom{***************}\cr\cr 
+\l\int\!d\ww\ 
v(\xx-\ww)\lft[
{\dpr^2\VV^{(k)}\over\dpr J_{\ww,\o,-\s} \dpr \f^+_{\xx,\s}}
+{\dpr \VV^{(k)}\over\dpr J_{\ww,\o,-\s}}
{\dpr \VV^{(k)}\over\dpr \f^+_{\xx,\o,\s}}\rgt](\h,J,0)\;.
\eea
Moreover the {\it Wick theorem} for Gaussian mean values gives
\bea
\lb{mas1b}
\int\!P(d\psi^{[k+1,N]})\; 
\ps_{\xx,\o,\s}^{[k+1,N]-} F\big(\ps^{[k+1,N]}\big)
\cr\cr
=\int\!d\uu\ 
g_\o^{[k+1,N]}(\xx-\uu)
\int\!P(d\psi^{[k+1,N]})\; 
{\dpr F\over \dpr \ps_{\uu,\o,\s}^+}
\big(\ps^{[k+1,N]}\big)
\eea
for $F$ any polynomial in the field. As direct application, we obtain
\bea
&&{\dpr \VV^{(k)}\over\dpr \f^+_{\xx,\o,\s}}(\h,J,\f)
=e^{-\VV^{(k)}(\h,J,\f)}{\dpr e^{\VV^{(k)}(\h,J,\f)}\over\dpr \f^+_{\xx,\o,\s}}
\cr\cr
&&=e^{-\VV^{(k)}(\h,J,\f)}\int\!P(d\psi^{[k+1,N]})\ 
\lft(\ps_{\xx,\o,\s}^{[k+1,N]-}+\h_{\xx,\o,\s}^-\rgt) e^{\VV^{(N)}(\ps+\h,J,\f)}
\cr\cr \lb{mas2}
&&=
\h_{\xx,\o}^- 
+\int\! d\uu\ 
g_\o^{[k+1,N]}(\xx-\uu) {\dpr \VV^{(k)}\over \dpr \h_{\uu,\o,\s}^+}(\h,J,\f)\;.
\eea
Another useful consequence is
(since $g_\o(0)=0$):
\bea
{\dpr \VV^{(k)}\over \dpr J_{\xx,\o,\s}}(\h,J,\f)
&=&
\h^+_{\xx,\o,\s}\h^-_{\xx,\o,\s}\lb{mas2b}
\cr\cr
&+&\int\!d\uu\ 
g_\o^{[k+1,N]}(\xx-\uu)\lft[
{\dpr \VV^{(k)}\over \dpr \h^-_{\uu,\o,\s}}\h^-_{\xx,\o,\s}+
\h^+_{\xx,\o,\s}{\dpr \VV^{(k)}\over \dpr \h^+_{\uu,\o,\s}}\rgt]
\cr\cr
&+&\int\!d\uu d\uu'\ 
g_\o^{[k+1,N]}(\xx-\uu)g_\o^{[k+1,N]}(\xx-\uu')
\cdot\cr\cr
&&\phantom{****}\cdot
\lft[
{\dpr^2 \VV^{(k)}\over \dpr \h^+_{\uu,\o,\s}\dpr \h^-_{\uu',\o,\s}}+
{\dpr\VV^{(k)}\over \dpr \h^+_{\uu,\o,\s}}
{\dpr\VV^{(k)}\over\dpr \h^-_{\uu',\o,\s}}\rgt]
\eea
We will use the following straightforward bounds, for $c_0,c_1,c_2>1$:
\bea\lb{dec}
|g_\o^{(h)}|_0\defi \sup_{\xx} |g_\o^{(h)}(\xx)|\le c_0 \g^{h}\;,
\cr\cr
|g_\o^{(h)}|_1\defi \int\!d\xx\;|g_\o^{(h)}(\xx)|\le c_1 \g^{-h}\;,
\cr\cr
\int\!d\xx\;|x_j||g_\o^{(h)}(\xx)|\le c_2 \g^{-2h}\;.
\eea
%

We start the improvement of the dimensional bounds by  considering 
$W^{(0;2)(k)}_{\s}$.  By symmetry we
have $W^{(1;0)(k)}_{-\s}(\ww)\=0$; hence
from \pref{mas1} and \pref{mas2}
we expand the two-points kernel as in Fig. \ref{q1}
\insertplot{190}{50}
{\ins{12pt}{28pt}{$\xx$}
\ins{53pt}{28pt}{$\yy$}

\ins{78pt}{31pt}{$=$}

\ins{102pt}{28pt}{$\xx$}
\ins{126pt}{16pt}{$\ww'$}
\ins{128pt}{57pt}{$\ww$}
\ins{170pt}{28pt}{$\yy$}

%

}%
{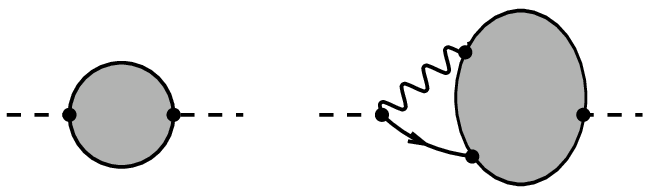}{\lb{q1}: Topological identity for $W^{(0;2)(k)}$}{0}
\\
\bea\lb{111}
W^{(0;2)(k)}_{\s}(\xx,\yy)
\cr\cr=
\l \int\!d\ww d\ww'\; 
v(\xx-\ww) g_\o^{[k+1,N]}(\xx-\ww')W^{(1;2)(k)}_{-\s;\s}(\ww;\ww',\yy)
\eea
so that, from the bound \pref{pc},  $\|W_{-\s;\s}^{(1;2)(k)}\|_k\le C$
given by \pref{pc}, and by the second of \pref{dec},
we obtain
\bea
\lb{111b}
\|W^{(0;2)(k)}_{\s}\|\le |\l|\cdot |v|_\io\cdot
\|W_{-\s;\s}^{(1;2)(k)}\|_k\cdot  
\sum_{j=k}^N |g^{(j)}_\o|_1
\cr\cr
\le  {c_1\over 1-\g^{-1}}C |v|_\io|\l|\g^{-k}
\le {1-\g^{-1}\over 4 c_1}C_1|\l|\g^{-k}\;.
\eea
which proves the first of \pref{hb}, since ${1-\g^{-1}\over c_1}<1$
($C_1$ is chosen so large to have such a factor because of later usage). 
Let us consider now $W^{(1;2)(k)}_{\s';\s}$, which
from  \pref{mas1} can be rewritten as in Fig. \ref{q2}
\insertplot{310}{140}
{\ins{12pt}{98pt}{$\xx$}
\ins{53pt}{98pt}{$\yy$}
\ins{23pt}{120pt}{$\zz$}

\ins{85pt}{105pt}{$-\d_{\s',\s}$}
\ins{120pt}{95pt}{$\zz=\xx=\yy$}

\ins{170pt}{102pt}{$=$}

\ins{203pt}{95pt}{$\xx$}
\ins{230pt}{86pt}{$\uu$}
\ins{265pt}{86pt}{$\yy$}
\ins{230pt}{130pt}{$\ww$}
\ins{265pt}{130pt}{$\zz$}
\ins{245pt}{143pt}{(a)}

\ins{25pt}{43pt}{$+$}

\ins{39pt}{7pt}{$\xx=\yy$}
\ins{39pt}{26pt}{$\ww$}
\ins{41pt}{57pt}{$\zz$}
\ins{65pt}{73pt}{(b)}

\ins{85pt}{43pt}{$+$}

\ins{123pt}{7pt}{$\xx$}
\ins{147pt}{7pt}{$\uu$}
\ins{183pt}{7pt}{$\yy$}
\ins{119pt}{26pt}{$\ww$}
\ins{121pt}{57pt}{$\zz$}
\ins{161pt}{57pt}{(c)}

\ins{190pt}{43pt}{$+\d_{\s',\s}$}

\ins{215pt}{27pt}{$\xx=\zz$}
\ins{247pt}{27pt}{$\uu$}
\ins{283pt}{27pt}{$\yy$}
\ins{251pt}{57pt}{(d)}
}%
{q2}{\lb{q2}: Topological identity for  $W^{(1;2)(k)}$}{0}
\*
\bd
\item{1.}
The graph (a)  in Fig.\ref{q2} is  given by:
\bea
W^{(1;2)(k)}_{(a)\s';\s}(\zz;\xx,\yy)
\cr\cr
\defi
\l\int\! d\ww d\uu\
v(\xx-\ww)g^{[k+1,N]}_\o(\xx-\uu)W^{(2;2)(k)}_{\s',-\s;\s}(\zz,\ww;\uu,\yy) 
\eea
From the bound \pref{pc}, $\|W^{(2;2)(k)}_{\s',-\s;\s}\|_k\le C^2
\g^{-k}$, we obtain
\bea
\|W^{(1;2)(k)}_{(a);\s';\s}\|_k
\le |\l|\cdot |v|_\io\cdot  
\|W^{(2;2)(k)}_{\s',-\s;\s}\|_k
\cdot
\sum_{j=k}^N |g^{(j)}_\o|_1
\le {C_1\over 4}|\l|\g^{-2k}
\eea
\item{2.} The graph (d) is given by
\bea
W^{(1;2)(k)}_{(d)\s';\s}(\zz;\xx,\yy)
\defi
\d_{\s,\s'}
\d(\xx-\zz)\int\! d\uu\ 
g_\o^{[k+1,N]}(\xx-\uu) W^{(0;2)(k)}_{\s}(\uu,\yy)
\eea
and using \pref{111b} we get 
\bea
\|W^{(1;2)(k)}_{(d)\s';\s}\|_k
\le
\d_{\s,\s'}\cdot \| W^{(0;2)(k)}_\s\|_k\cdot
\sum_{j=k}^N |g^{(j)}_\o|_1
\cr\cr  \le \| W^{(0;2)(k)}_\s\|_k
\cdot {c_1\over 1-\g^{-1}} \g^{-k}
\le 
{C_1\over 4}|\l|\g^{-2k}
\eea
\ed
In order to obtain an improved bound also for the 
graphs (b) and (c) of Fig. \ref{q2}, we need to further 
expand $W^{(2;0)(k)}_{\s';-\s}$. Using \pref{mas2b}, we find
\bea\lb{63}
W^{(2;0)(k)}_{\s',-\s}(\zz,\ww)
 \cr\cr=
\int\! d\uu'd\uu\ 
g_{\o}^{[k+1,N]}(\ww-\uu)g_{\o}^{[k+1,N]}(\ww-\uu')
W^{(1;2)(k)}_{\s';-\s}(\zz;\uu',\uu)
\eea
and then, replacing the expansion 
for $W^{(1;2)(k)}_{\s';-\s}(\zz;\uu',\uu)$
in the graph \pref{63} we find
for (b) what is depicted in Fig.\ref{q3}:
\insertplot{270}{160}
{\ins{3pt}{125pt}{$\xx$}
\ins{25pt}{125pt}{$\ww$}
\ins{50pt}{153pt}{$\uu'$}
\ins{50pt}{114pt}{$\uu$}
\ins{78pt}{125pt}{$\zz$}
\ins{14pt}{153pt}{(b)}

\ins{116pt}{132pt}{$=$}

\ins{138pt}{124pt}{$\xx$}
\ins{158pt}{124pt}{$\ww$}
\ins{170pt}{146pt}{$\uu'$}
\ins{196pt}{128pt}{$\zz'$}
\ins{217pt}{116pt}{$\uu$}
\ins{217pt}{156pt}{$\ww'$}
\ins{227pt}{124pt}{$\zz$}
\ins{148pt}{159pt}{(b1)}

\ins{-18pt}{84pt}{$+\ \d_{\s',-\s}$}

\ins{13pt}{73pt}{$\xx$}
\ins{33pt}{73pt}{$\ww$}
\ins{82pt}{73pt}{$\zz$}
\ins{28pt}{103pt}{(b2)}

\ins{112pt}{84pt}{$+$}

\ins{124pt}{73pt}{$\xx$}
\ins{144pt}{73pt}{$\ww$}
\ins{192pt}{73pt}{$\uu$}
\ins{202pt}{76pt}{$\zz'$}
\ins{235pt}{73pt}{$\zz$}
\ins{195pt}{103pt}{(b3)}

\ins{-18pt}{35pt}{$+\ \d_{\s',-\s}$}

\ins{12pt}{23pt}{$\xx$}
\ins{32pt}{23pt}{$\ww$}
\ins{82pt}{23pt}{$\zz$}
\ins{36pt}{14pt}{$\ww'$}
\ins{74pt}{14pt}{$\zz'$}
\ins{25pt}{53pt}{(b4)}

\ins{112pt}{33pt}{$+$}

\ins{122pt}{23pt}{$\xx$}
\ins{142pt}{23pt}{$\ww$}
\ins{192pt}{23pt}{$\uu$}
\ins{146pt}{14pt}{$\ww'$}
\ins{184pt}{14pt}{$\zz'$}
\ins{200pt}{26pt}{$\uu'$}
\ins{234pt}{23pt}{$\zz$}
\ins{193pt}{53pt}{(b5)}



}%
{q3}{\lb{q3}: Graphical representation of graph (b) in Fig.\ref{q2} }{0}
\\
\bd
\item{3.}
We now consider (b1) of Fig.\ref{q3}. 
\bea
W^{(1;2)(k)}_{(b1)\s';\s}(\zz;\xx,\yy)
\defi
\l\d(\xx-\yy)\int\! d\ww d\uu' d\zz'\ 
v(\xx-\ww) v(\uu'-\zz')\cr
\int\! d\uu  d\ww'\ g_{\o}^{[k+1,N]}(\ww-\uu)g_{\o}^{[k+1,N]}(\ww-\uu')
g_{\o}^{[k+1,N]}(\uu'-\ww')
\cdot\cr\cr\cdot W^{(2;2)(k)}_{\s',\s;-\s}(\zz,\zz';\ww',\uu)
\eea
In order to obtain bound uniform in $N-k$, it is convenient to 
decompose the three propagators $g_{\o}g_{\o}g_{\o}$ into scales,
$\sum_{j,i,i'=k}^N g^{(j)}_{\o}g^{(i)}_{\o} g^{(i')}_{\o}$
and then, for any realization of $j,i,i'$, to take 
the $|\cdot|_1$ norm on the two propagator on the higher scales,
and the $|\cdot|_\io$ norm  on the propagator with the lowest one. 
In this way we obtain:
\bea\lb{27}
\|W^{(1;2)(k)}_{(b1)\s';\s}\|\le 
|\l| \cdot|v|_\io
\cdot |v|_1
\cdot\|W^{(2;2)(k)}_{\s',\s,-\s}\|_k
\cdot\cr\cr\cdot
3!\sum_{j=k}^N\sum_{i=k}^j\sum_{i'=k}^i 
|g^{(j)}_{\o}|_1 |g^{(i)}_{\o}|_1 |g^{(i')}_{\o}|_\io    
\le {C_1\over 20}
|\l| \g^{-2k}
\eea
where, in the last inequality, we have taken $|\l|$  small enough, and
we have used that $\sum_{j=k}^N \g^{-j}(j-k)\le C \sum_{j=k}^N \g^{-j}\g^{(j-k)/2}
\le C'\g^{-k}$.
\item{4.} The expression for (b2) is:
\bea
W^{(1;2)(k)}_{(b2)\s';\s}(\zz;\xx,\yy)
\cr\cr
\defi
\l \d_{\s',-\s}\d(\xx-\yy)\int\!d\ww\ 
v(\xx-\ww)
\lft[g^{[k+1,N]}_{-\o}(\ww-\zz)\rgt]^2
\eea
For $\kk^*=(-k_0,k)$,
it holds
$\hat g_\o^{[k+1,N]}(\kk)=-i\o \hat g_\o^{[k+1,N]}(\kk^*)$ hence
\be
\int\!  d\uu\ 
\lft[g^{[k+1,N]}_{-\o}(\uu)\rgt]^2=0\;.\lb{mas3}
\ee
Since
\be \lb{idb}
v(\xx-\ww)=v(\xx-\zz)+
\sum_{j=0,1} (z_j-w_j) \int_0^1\!\!d \t\ 
\big(\dpr_j v\big)\big(\xx-\zz+\t(\zz-\ww)\big)
\ee
we can write
\bea 
W^{(1;2)(k)}_{(b2)\s';\s}(\zz;\xx,\yy)
=\l \d_{\s',-\s}\d(\xx-\yy) v(\xx-\zz) 
\int\!d\ww\ 
\lft[g^{[k+1,N]}_{-\o}(\ww)\rgt]^2
\cr\cr
+\l \d_{\s',-\s}\d(\xx-\yy)
\cdot\cr\cr\cdot
\sum_{j=0,1} \int_0^1\!\!d \t\ \int\! d\ww\ 
\big(\dpr_j v\big)\big(\xx-\zz+\t(\zz-\ww)\big)
(z_j-w_j)g^{[k+1,N]}_{\o}(\ww-\zz)\nn
\eea
and the first addend is vanishing because of \pref{mas3}.
Hence, using the third of \pref{dec}, 
\bea
\|W^{(1;2)(k)}_{(b2)\s';\s}\|_k
\cr\cr\le |\l| 
\sum_{j=0,1} \int_0^1\!\!d \t\ 
\int\! d\ww d\xx\ 
\lft|\big(\dpr_j v\big)\big(\xx-\zz-\t\ww\big)
w_j \lft[g^{[k+1,N]}_{-\o}(\ww)\rgt]^2\rgt|
\cr\cr
\le
4|\l| \int\!d\xx\ 
 \big| (\dpr_j v) (\xx)\big|
\sum_{i=k}^N\sum_{j=k}^i
|g^{(j)}_{-\o}|_\io
\cdot\cr\cr\cdot
\int\!d\ww\
|w_j||g^{(i)}_{\o}(\ww)|
\le |\l| {C_1\over 20}\g^{-k}
\eea
\item{5.}The expression for (b3) is:
\bea
W^{(1;2)(k)}_{(b2)\s';\s}(\zz;\xx,\yy)
\defi
\l \d_{\s',-\s}\d(\xx-\yy)\int\!d\ww\ 
v(\xx-\ww)
\cdot\cr\cr\cdot
\l\int\! d\zz'\ 
\lft[g^{[k+1,N]}_{-\o}(\ww-\zz')\rgt]^2
v(\uu-\zz')
W^{(2;0)(k)}_{\s,\s'}(\zz',\zz)
\eea
The improved bound for (b3) is obtained in the same way as for (b1).
\bea
\|W^{(1;2)(k)}_{(b3)\s';\s}\|_k
\le {C_1\over 20}|\l| \g^{-k}\;.
\eea
\item{6.}
It is convenient to further expand (b4) using the identity \pref{111},
which, in the case at hand, is depicted in Fig \ref{q3b}.
\insertplot{230}{50}
{\ins{99pt}{32pt}{$=$}

\ins{2pt}{23pt}{$\xx$}
\ins{22pt}{23pt}{$\ww$}
\ins{72pt}{23pt}{$\zz$}
\ins{26pt}{14pt}{$\ww'$}
\ins{64pt}{14pt}{$\zz'$}

\ins{112pt}{23pt}{$\xx$}
\ins{132pt}{23pt}{$\ww$}
\ins{202pt}{23pt}{$\zz$}
\ins{167pt}{1pt}{$\uu$}
\ins{167pt}{34pt}{$\uu'$}
\ins{135pt}{11pt}{$\ww'$}
\ins{194pt}{15pt}{$\zz'$}
}%
{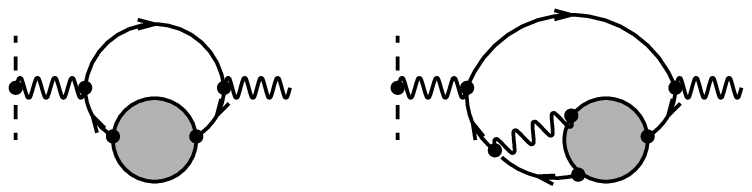}{\lb{q3b}: Equivalent expressions for (b4)}{0}
\\
Thereby,  explicit expression for (b4) is
\bea
W^{(1;2)(k)}_{(b4)\s';\s}(\zz;\xx,\yy)
\defi
\d_{\s',-\s}
\l^2\int\! d\zz' d\ww\  
v(\xx-\ww) g_{\o}(\ww-\zz)
\cdot\cr \cr\cdot
\int\! d\ww'd\uu'd\uu\ 
g_{\o}^{[k+1,N]}(\ww-\ww')g_{\o}^{[k+1,N]}(\ww'-\uu)
v(\ww'-\uu')
\cdot\cr \cr\cdot
W^{(1;2)(k)}_{-\s;\s}(\uu';\uu,\zz') g^{[k+1,N]}_{\o}(\zz'-\zz)
\eea
As in the previous cases, it is convenient
first to decompose the propagators 
$g_{\o}(\ww-\zz)g_{\o}(\ww-\ww')g_{\o}(\ww'-\uu)$ into scales,
$\sum_{j,i,i'=k}^N g^{(j)}_{\o}g^{(i)}_{\o}g^{(i')}_{\o}$ 
and then, for any realization of $j,i,i'$,
to bound with $|\cdot|_1$ norm the two propagators on highest scale, and
with $|\cdot|_\io$ norm the one on lower scale. 
Finally, for $|\l|$ small enough, we have:
\bea
\|W^{(1;2)(k)}_{(b4);\o';\o}\|_k\le 
\d_{\s',-\s}
|\l|^2\cdot|v|_1\cdot|v|_\io\cdot\|W^{(1;2)(k)}_{-\s;\s}\|_k\cdot
|g_{\o}|_1
\cdot\cr \cr\cdot
3!\sum_{j=k}^N\sum_{i=k}^j\sum_{i'=k}^i
|g_\o^{(j)}|_1\;|g_\o^{(i)}|_1\;|g_\o^{(i')}|_\io
\le {C_1\over20} |\l| \g^{-2k}
\eea
\item{7.}
Similar arguments can be 
used to bound also the graph $(b5)$.
\ed
Finally, it is also clear that a bound for (c) of Fig. \ref{q2} 
can be found along the
same lines discussed for (b) of the same figure. 
We have so proved, therefore
\be\lb{ab}
\|W^{(1;2)(k)}_{\s';\s}\|_k\le {C_1\over C_2}\g^{-k}
\ee
where, for later purposes, $C_1$ is chosen large enough so that in
\pref{ab} 
$C_2=1+2|v|_\io\lft(1+|g|_1\cdot \|W^{(0;2)(k)}_{\s}\|_k\rgt)$. 
Clearly \pref{ab} implies the second of \pref{hb}.

Finally
from  \pref{mas1} we obtain the identity in Fig. \ref{q7}.
\insertplot{280}{170}
{\ins{12pt}{148pt}{$\xx$}
\ins{43pt}{135pt}{$\yy$}
\ins{47pt}{175pt}{$\yy'$}
\ins{52pt}{149pt}{$\xx'$}

\ins{80pt}{155pt}{$-\d_{-\s,\s'}$}

\ins{105pt}{136pt}{$\xx=\yy$}
\ins{127pt}{168pt}{$\xx'=\yy'$}

\ins{179pt}{152pt}{$=$}

\ins{188pt}{135pt}{$\xx=\yy$}
\ins{221pt}{163pt}{$\ww$}
\ins{257pt}{178pt}{$\yy'$}
\ins{257pt}{143pt}{$\xx'$}
\ins{201pt}{163pt}{(a)}

\ins{24pt}{95pt}{$+$}

\ins{60pt}{85pt}{$\xx$}
\ins{85pt}{75pt}{$\uu$}
\ins{80pt}{112pt}{$\ww$}
\ins{117pt}{75pt}{$\yy$}
\ins{111pt}{95pt}{$\xx'$}
\ins{112pt}{126pt}{$\yy'$}
\ins{61pt}{123pt}{(b)}

\ins{165pt}{93pt}{$+$}

\ins{194pt}{86pt}{$\xx$}
\ins{221pt}{120pt}{$\ww$}
\ins{221pt}{74pt}{$\uu$}
\ins{257pt}{122pt}{$\yy'$}
\ins{260pt}{107pt}{$\xx'$}
\ins{257pt}{78pt}{$\yy$}
\ins{201pt}{123pt}{(c)}

\ins{23pt}{35pt}{$+\d_{\s,\s'}$}

\ins{48pt}{17pt}{$\xx=\yy'$}
\ins{86pt}{42pt}{$\ww$}
\ins{111pt}{56pt}{$\yy$}
\ins{112pt}{25pt}{$\xx'$}
\ins{61pt}{43pt}{(d)}

\ins{165pt}{35pt}{$+\d_{\s,\s'}$}

\ins{205pt}{17pt}{$\xx$}
\ins{228pt}{44pt}{$\ww$}
\ins{229pt}{6pt}{$\uu$}
\ins{255pt}{55pt}{$\yy$}
\ins{253pt}{25pt}{$\xx'$}
\ins{258pt}{5pt}{$\yy'$}
\ins{201pt}{44pt}{(e)}
}%
{q7}{\lb{q7}:Graphical representation of $W^{(0;4)(k)}$. 
The dark bubble represents $W^{(1;2)(k)}_{\s;\s'}-\d_{\s,\s'}\d_2$. }{0}
\\
Therefore the bound for the sum of the graphs (a), (b), (d), and (e) is 
\bea
|\l|\cdot|v|_1\cdot 
\|W^{(1;2)(k)}_{\s;\s'}-\d_{\s,\s'}\d_2\|_k
\lft(1+|g|_1\cdot \|W^{(0;2)(k)}_{\s}\|_k\rgt)
\le {C_1\over 2}\g^{-k}\;.
\eea
Indeed, the last inequality follows from
the just proved, improved bound
$\|W^{(1;2)(k)}_{\s;\s'}-\d_{\s,\s'}\d_2\|_k\le {C_1\over C_2}|\l|\g^{-k}$.
Finally, the graph (c) is 
\bea
W^{(0;4)}_{(a),\o;\s,\s'}(\xx,\yy,\xx',\yy')
\cr\cr
\defi
\l\int\!d\ww d\uu\ 
v(\xx-\ww)g_\o^{[k+1,N]}(\xx-\uu)W^{(1;4)}_{-\s;\s,\s'}(\ww;\uu,\yy, \xx',\yy')
\eea
Using \pref{pc},  $\|W^{(1;4)}_{-\s;\s,\s'}\|\le
C|\l|\g^{-k}$ and 
\bea
\|W^{(0;4)}_{(a),\o;\s,\s'}\|_k
\le
|\l| \cdot |v|_\io\cdot |g_\o|_1
\cdot\|W^{(1;4)}_{-\s;\s,\s'}\|_k
\le {C_1\over 2} |\l| \g^{-2k}
\eea
From this the third of \pref{hb} follows and the theorem is proved. \qed

\section{Schwinger functions}\lb{s4}
The multiscale integration of \pref{gf1}, when $\f\not=0$, is obtained by a slight
modification of the one presented in \S 2. 
In particular $\VV^{(k)}(\psi^{(\le k)},\phi,J)$
is given by an expression similar to \pref{ker}, sum of monomials
in $\psi^{(\le k)},J$ and $\phi$. We define $\LL=0$ on the kernels of the monomials
containing at least a $\phi$ except when the monomial is
$\f^+_{\xx,\o,\s} \ps^{-(\le k+1)}_{\yy,\o,\s}$ 
or $\ps^{+(\le k+1)}_{\yy,\o,\s} \f^-_{\xx,\o,\s} $;
in such a case the kernel is $\hg_\o(\kk)\hW^{(0;2)(k)}_{\s}(\kk)$ and
we define, for $0\le k\le N$,
\be\lb{7t}
\LL \lft[\hg_\o(\kk)\hW^{(0;2)(k)}_{\s}(\kk)\rgt]
\defi\hg_\o(\kk)\hW^{(0;2)(k)}_{\s}(\kk)
\ee
while, for $k<0$, $\LL\=0$.
Correspondingly , for $k>0$ we define
\be\lb{8t}
\g^{-k}\tilde \n_{k,\o,\s}(\xx,\yy)\defi
\int\!d\zz\;
g_{\o}(\xx-\zz) W^{(0;2)(k)}_{\s}(\zz,\yy).
\ee
and using \pref{111b} we obtain 
$\|\tilde \n_{k,\o,\s}\|_k\le C_1|\l|\g^{-k} $; 
while for $k<0$ we set $\tilde \n_{k,\o,\s}(\xx,\yy)\=0$, 
because of the fact that $\hW_{\s}^{0;2(k)}(0)=0$ by symmetries,
and then there is an automatic dimensional gain:
\be\lb{adg}
\hg_\o(\kk)\hW_{\s}^{(0;2)(k)}(\kk)=
\hg_\o(\kk)\lft[\hW_{\s}^{(0;2)(k)}(\kk)-\hW_{\s}^{(0;2)(k)}(0)\rgt]
\ee
Let $\tilde
\e_k$ be larger than $\e_k$ and  
$\max_{\o,\s}\max_{h:k\le h\le N}\|\tilde  \n_{k,\o,\s}\|_k$.
The 2-points Schwinger function  is given by
\be
S_{N;\o;\s}(\xx,\yy)=
\sum_{h\le N} g^{(h)}_\o(\xx-\yy)+
\sum_{n=0}^\io\sum_{j\le N}
\sum_{\t\in\bar\TT_{j,n}^{2,0}} \sum_{\bP\in \PP
\atop |P_{v_0}|=2} S_\t(\xx,\yy)\;, \lb{2.60}\ee
where $\bar\TT_{h,n}^{n^\f,n^J}$ is the set of trees with
$n$ endpoints, $n^\f$ special endpoints of type $\f$, $n^J$
endpoints of type $J$ and first vertex scale $j$; $n^J_v,n^\phi_v$
are the number of fields of type $J,\phi$ associated to end-points
following $v$. If $h$ is the
first nontrivial vertex $u$ of $\t$, and $h_1$ and $h_2$ are the scale of
the two endpoints of type $\f$, we have
\bea\lb{sf}
|S_\t(\xx,\yy)|\le \tilde C_q
(c\tilde \e_h)^{n-2}
\g^{j-h_1-h_2}
\prod_{v\ {\rm not}\ {\rm e. p.}}\g^{-({|P_v|\over 2}-2+z_v)}
\cr\cr
{\g^{2h}\over 1+[\g^h|\xx-\yy|]^{q\over 2}}
\eea
Indeed, \pref{sf} is the same of \pref{38}, for $|P_{v_0}|=2$,
$n^J_{v_0}=0$, times some factors more. 
\bd
\item{1.}
The presence, with respect to the graphical expansion of
the kernels, of two external propagators,  $g^{(h_1)}_\o$ and
$g_\o^{(h_2)}$, causes the factor  $\g^{-h_1-h_2}$.
\item{2.}
Before performing the bounds as for the kernels, it is possible to
extract from the bound on the propagator \pref{17}
a factor $b_h=\big(1+\lft(\g^h|\xx-\yy|\rgt)^{q\over 2}\big)^{-1}$:
the product   of $b_h$ for each of the propagators
of the graph that are not involved into the Gram determinant \pref{gd}
can be bounded with the factor $\big[ 1+[\g^h|\xx-\yy|]^{q\over 2}\big]^{-1}$ 
in \pref{sf} at the price of a constant $C^n$. 
\item{3.}
The bounds for the kernels can be straightforwardly  modified also for
obtaining the factor $\g^{2h}$: it is the effect of the missed
integration in the variable $\xx-\yy$, that causes 
the replacement of $|\cdot|_1$-norm with the $|\cdot|_\io$-norm 
of a propagator  $g_\o$; this occurs  in correspondence of $v$, 
the vertex with highest scale, $h$, in which the two
special endpoints of type $\f$ are connected. 
\ed
It is convenient to call $|P_v|=n^\ps_v+n^\f_v$. We have that $z_v$ is
the same of \pref{38}, with a further case in which it is not zero:
if $h_v<0$ and $n^\ps_v=n^\f_v=1$, then $z_v=1$. This is because the
automatic dimensional gain depicted in \pref{adg}.

Along the tree $\t$, we consider three paths: $\CC_1$ and $C_2$, connecting the
endpoint of type $\f$ on scale $h_1$ and the one on scale $h_2$
respectively with $v_0$; and $\CC$ connecting
$u$ with $v_0$. For $j=1,2$, we find $
\g^{-h_j}=\g^{-j}\prod_{v\in \CC_j}\g^{-1}$ and 
$\g^{-j}=\g^{-h}\prod_{v\in \CC}\g$.
These identities, replaced in \pref{sf}, gives:
\bea\lb{sf2}
|S_\t(\xx,\yy)|\le \tilde C_q
(c\tilde \e_h)^{n-2}
{\g^{h}\over 1+[\g^h|\xx-\yy|]^{q\over 2}}
\cr\cr
\lft(\prod_{v\ {\rm not}\ {\rm e. p.}}^{v\not\in
  \CC}\g^{-({n^\ps_v\over2} +{3n^\f_v\over 2}-2+z_v)}\rgt)
\prod_{v\in\CC}\g^{-{n^\ps_v\over2}} 
\eea
and ${n^\ps_v\over2} +{3n^\f_v\over 2}-2+z_v>0$, as well as
${n^\ps_v\over2}>0$ for $v\in \CC$: we can perform the summation on the
trees, keeping fixed the scale $h$.
\be
|S_{N;\o;\s}(\xx,\yy)-g^{(\le N)}_\o(\xx,\yy)|\le C |\l| \sum_{h\le N} 
{\g^h\over 1+[\g^h|\xx-\yy|]^{q\over 2}}\le C |\l|{1\over |\xx-\yy|}
\ee
Finally, we want to study the difference $S_{N;\o;\s}(\xx,\yy)-
S_{\o;\s}(\xx,\yy)$ for $\xx-\yy\neq 0$.
\bea
S_{N;\o;\s}(\xx,\yy)-S_{\o;\s}(\xx,\yy)=
\sum_{h\le N} g^{(h)}_\o(\xx-\yy)
\cr\cr+
\sum_{n=0}^\io\sum_{j=-\io}^{+\io}
\sum_{\t\in\bar\TT_{j,n}^{2,0}} \sum_{\bP\in \PP
\atop |P_{v_0}|=2} D_\t(\xx,\yy)\;,
\eea
In such a tree expansion, $D_\t(\xx,\yy)$ is not zero only in two cases: 
either $\t$ has at least one vertex $v^*$ on scale $h^*>N$; or $\t$ has
vertices scales $\le N$, but has an endpoint which, in turn, has
tree expansion with at least one vertex  $v^*$ on scale $h^*>N$. 
If $\t$ is of the former type, fixed $\th$, we have
\bea
|D_\t(\xx,\yy)|\le \tilde C_q
(c\tilde \e_h)^{n-2}
{\g^{h}\over 1+[\g^h|\xx-\yy|]^{q\over 2}}
\cr\cr
\g^{-\th(h^*-h)}
\lft(\prod_{v\ {\rm not}\ {\rm e. p.}}^{v\not\in
  \CC}\g^{-({n^\ps_v\over2} +{3n^\f_v\over 2}-2+z_v-\th)}\rgt)
\prod_{v\in\CC}\g^{-{n^\ps_v\over2}} 
\eea
It $\t$ is of the latter type, we still have the above bound, by
induction on the subtrees in which the endpoints can be expanded:
indeed, in the analysis of the previous section it is clear that if
the fermion propagator is constrained to be on scale $>N$, bounds
\pref{hb} are still true, with a more factor $\g^{-\th(h^*-k)}$, which,
together to a factor $\g^{-\th(k-h)}$ gives the wanted $\g^{-\th(h^*-h)}$.

For $\th>0$ and but small enough, we still have 
${n^\ps_v\over2} +{3n^\f_v\over 2}-2+z_v-\th>0$. This means that 
we can perform the summation on the
trees, keeping fixed the scale $k$. As $\g^{-\th(h^*-h)}\le
\g^{-\th(N-h)}$,

\bea\lb{87}
|S_{N;\o;\s}(\xx,\yy)-S_{\o;\s}(\xx,\yy)|\le C \g^{-\th N}\sum_{h=-\io}^{+\io}
{\g^{(1+\th)h}\over 1+[\g^h|\xx-\yy|]^{q\over 2}}
\cr\cr\le C {1\over \g^{\th N}|\xx-\yy|^{1+\th}}
\eea
\section{Ward Identities}
Let us consider the 2-point Schwinger function with one density
insertion:
\be 
\hG_{N;\o,\s';\s}(\pp;\kk)={\partial^3 \WW\over \partial \hJ_{\pp,\o,\s'} 
\partial\hp^+_{\kk+\pp,\o,\s}\partial\hp^-_{\kk,\o,\s}}(0,0)
\ee
In the generating functional \pref{gf1}, we perform 
the phase-chiral transformation
\be
\hp^\e_{\kk,\o,\s}\to 
\hp^\e_{\kk,\o,\s}+
\e\int\!{d \pp\over (2\p)^2}\ha_{\pp,\o,\s}\hp^\e_{\kk+\e\pp,\o,\s}
\ee
and obtain the identities:
\bea
D_{\o}(\pp)\hG_{N;\o,\s';\s}(\pp;\kk)
\cr\cr
=\d_{\s,\s'} 
\lft[\hS_{N;\o,\s}(\kk)-\hS_{N;\o,\s}(\kk+\pp) \rgt]
+\D_{\o,\s';\s}(\pp;\kk)
\eea
where $\D_{\o;\s',\s}(\pp;\kk)$ is a correction term caused by
the presence of the cutoff:
$$\D_{\o;\s',\s}(\pp;\kk)=
\int\!{d\qq\over (2\p)^2}\;
C_{N;\o}(\qq+\pp,\qq)\la \hp^+_{\pp+\qq,\o,\s'}\hp^-_{\pp,\o,\s'}
\hp^-_{\kk+\pp,\o,\s}\hp^+_{\kk,\o,\s}\ra$$
for
$$C_{N;\o}(\kk+\pp,\kk)\defi
D_\o(\kk+\pp)\lft[1-\c_{N}^{-1}(\kk+\pp)\rgt]
-D_\o(\kk)\lft[1-\c_{N}^{-1}(\kk)\rgt]$$
The rest $\D_{\o;\s',\s}(\pp;\kk)$ does not vanish in the
limit of removed cutoff, but rather it causes the {\it anomaly} of the
Ward Identities.
\begin{theorem}\lb{th2}
There exists $\e_0>0$  such that, for $|\l|\le\e_0$  and in the limit
of removed cutoff,
\bea\lb{9}
\hG_{\o,\s';\s}(\pp;\kk)
={a(\pp)+\s\s'\bar a(\pp)\over 2}
\lft[\hS_{\o,\s}(\kk)-\hS_{\o,\s}(\kk)\rgt]
\eea
for 
$$
a(\pp)={1\over D_{\o}(\pp)-{\l\over 2\p}\hv(\pp)D_{-\o}(\pp)}\quad\quad
\bar a_N(\pp)={1\over D_{\o}(\pp)+ {\l\over 2\p}\hv(\pp)D_{-\o}(\pp)}
$$
\end{theorem}
The proof is a consequence of the two following lemmas.
\begin{lemma} For $|\l|$ small enough and $\pp,\kk,\pp-\kk\neq 0$,
the limit of removed cutoff of $\hG_{\o,\s';\s}(\pp;\kk)$ exist and is finite.
\end{lemma}
{\bf\0Proof.}
We can write
\be
\hG_{\s';\s}(\pp;\kk) = \sum_{n=0}^\io
\sum_{j\le N}\sum_{\t\in\TT_{j,n}^{2,1}} \sum_{\bP\in \PP
\atop |P_{v_0}|=2} \hG_\t(\pp;\kk)\;, \lb{2.60b}\ee
with an obvious definition of $\hG_\t(\pp,\kk)$.
We define $h_\pp= \min\{j: f_j(\pp)\not= 0\}$
and suppose that $\pp$, $\kk$, $\pp-\kk$ are all different from
$0$. It follows that, given $\t$, if $h_-$ and $h_+$ are the scale
indices of the $\psi$ fields belonging to the endpoints associated
with $\f^+$ and $\f^-$, while $h_J$ denotes the scale of the
endpoint of type $J$, $\hG_\t(\pp;\kk)$ can be different
from $0$ only if $h_-=h_\kk, h_\kk+1$, $h_+=h_{\kk-\pp},
h_{\kk-\pp} +1$ and $h_J\ge h_\pp-\log_\g2$. Moreover, if
$\TT_{j_0,n}^{\pp,\kk}$ denotes the set of trees satisfying the
previous conditions and $\t\in \TT_{j_0,n}^{\pp,\kk}$,
$|\hG_\t(\pp;\kk)|$ can be bounded by $\int\! d\zz d\xx\;
|G_\t(\zz; \xx, \yy)|$. We get
\bea
&&|\hG^{(1;2)}_{\s';\s}(\pp;\kk)| \le C \g^{-h_\kk}
\g^{-h_{\kk-\pp}}\;\cdot\cr\cr
&& \cdot \sum_{n=0}^\io \sum_{j\le N}
\sum_{\t\in\TT_{j_0,n}^{\pp,\kk}} \sum_{\bP\in \PP \atop
|P_{v_0}|=2} (C|\l|)^n \prod_{\rm v\ not\ e.p} \g^{-d_v} \;.
\lb{2.61}\eea
where $d_v={|P_v|\over 2}-2+z_v+n^\phi_v$.

Given $\t\in
\TT_{j_0,n}^{\pp,\kk}$, let $v^*_0$ the higher vertex preceding all
three special endpoints and $v^*_1\ge v^*_0$ the higher vertex
preceding either the two endpoints of type $\f$ 
or one endpoint of type $\f$ and the
endpoint of type $J$. 
We have $d_v>0$, 
except for a finite number of vertices belonging to the path $\CC^*$
connecting $v_1^*$ with $v^*_0$, where $d_v=0$:
\bd
\item{a)} the vertices with  $|P_v|=4$ and $n_v^J=0$;
since there is a momentum $\kk$ flowing inside the 
corresponding cluster and $\kk-\pp$ flowing outside, 
by conservation of the momenta the scale label 
of both of the other $\psi$ fields - and hence
also the scale label of such vertices - cannot be 
less than $\log_\g (|\pp|/2)$; 
\item{b)} the vertices with $|P_v|=2$ and $n_v^J=1$;
with a momentum $\pp$ flowing inside the cluster
and either a momentum $\kk$ flowing inside or $\kk-\pp$
flowing outside, the scale label of such vertices
cannot be less than $\min\{h_+,h_-\}-1$.
\ed
Accordingly, the number of the vertices depicted
in the above list is  not larger  than 
$\min\{|h_\kk-h_\pp|,|h_{\kk-\pp}-h_\pp|\}+2-\log_\g2$.
Thus we can replace in \pref{2.61} the rough bound:
$$
\prod_{\rm v\ not\ e.p} \g^{-d_v} \le C \g^{|h_\kk-h_\pp|}\g^{|h_{\kk-\pp}-h_\pp|}
\prod_{\rm v\ not\ e.p} \g^{-d_v-r_v}
$$
with $r_v=1$ for $v:d_v=0$ and $r_v=0$ otherwise.
Finally, we can perform the sums over the scale and $P_v$ labels of
$\t$, obtaining:
\be
|\hG_{\s';\s'}(\pp;\kk)| \le C \g^{-h_\kk}
\g^{-h_{\kk-\pp}}\g^{|h_\kk-h_\pp|}\g^{|h_{\kk-\pp}-h_\pp|}
\;.
\lb{2.62}\ee
This completes the proof. \qed

\begin{lemma}
There exist a finite $\n_N$ such that it is possible to decompose
\bea
\D^{(1;2)}_{\o;\s',\s}(\pp;\kk)-\n_N
\hv(\pp)D_{-\o}(\pp)\hG_{N;\o;-\s',\s}(\pp;\kk)
\cr\cr
=\sum_{\bar\o}D_{\bar\o}(\pp)\hR^{(1;2)}_{N;\bar\o,\o;\s',\s}(\pp;\kk)
\eea
where $\hR^{(1;2)}_{N;\bar\o,\o;\s',\s}$ is such that,
for fixed $\kk$ and $\pp$, it holds
\be\lb{89}
\lim_{N\to\io}\hR^{(1;2)}_{N;\bar\o,\o;\s',\s}(\pp;\kk)=0
\ee
Furthermore, 
$\lim_{N\to\io}\n_N={\l\over 2\pi}$.
\end{lemma}
{\bf\0Proof.}
It is convenient to write the rest $\hR^{(1;2)}_{\bar\o,\o,\s';\s}$ as
\be\lb{der}
\sum_{\o'}D_{\o'}(\qq)\hR^{(1;2)}_{\bar\o,\o,\s';\s}(\qq;\kk)
={\partial^3\WW_{\D}
\over\dpr \ha_{\qq,\o,-\s'}\dpr \hf^+_{\kk-\qq,\o,\s}
\dpr\hf^-_{\kk,\o,\s}} (0,0)
\ee
where we have introduced the new generating functional 
$\WW_{\D} (\a,\f)$ defined such that:
\bea\lb{96}
e^{\WW_{\D} (\a,\f)}=
\int\!P(d\psi^{\le N})\;e^{-\VV^{(N)}_\D(\psi^{(\le N)},\a,\f)}
\cr\cr
\defi
\int\!P(d\psi^{\le N})\;\exp\lft\{
-\l V(\psi^{(\le N)})+[T_{0}-\n_NT_{-}](\ps^{(\le N)},\a)\rgt\}
\cr\cr
\cdot \exp\lft\{\sum_{\o,\s}\int\!d\zz\;
\lft(\psi^{(\le N)+}_{\zz,\o,\s}\f^-_{\zz,\o,\s}+\f^+_{\zz,\o,\s}\psi^{(\le N)-}_{\zz,\o,\s}\rgt)\rgt\}
\eea
with
\bea\label{vvn}
&&T_{0}(\psi,\a)=
\sum_{\o,\s}\int\!{d\pp\; d\qq\over (2\p)^4}\ 
\bar\chi_N(\pp) C_{N;\o}(\qq+\pp,\qq)\ha_{\pp,\o,\s}
\hp^+_{\qq+\pp,\o,\s}\hp^-_{\qq,\o,\s}
\cr\cr
&&T_{-}(\psi,\a)=
\sum_{\o,\s}\int\!{d\pp\;d\qq\over (2\p)^4}\ 
\bar\chi_N(\pp) \hv(\pp)D_{-\o}(\pp)\ha_{\pp,\o,\s}
\hp^+_{\qq+\pp,\o,-\s}\hp^-_{\qq,\o,-\s}
\eea
We remark that the  presence of the cutoff function 
$\bar\c_N(\pp)\defi \sum_{j=-N}^N \haf_j(\pp)$ is immaterial for
\pref{der}, since $\pp$ is finite and nonzero.  But it is essential
for the multiscale integration, because it simplify the discussion of
the {\it tadpoles.}

A crucial role in the following analysis is played by the functions
\bea
\hU^{(i,j)}_\o(\qq+\pp,\qq)
\defi
\bar\c_N(\pp)C_{N;\o}(\qq+\pp,\qq)
\hg^{(i)}_\o(\qq+\pp) \hg^{(j)}_\o(\qq)
\cr\cr
\hQ^{(N,i)}_\o(\qq+\pp,\qq)
\defi
\bar\c_N(\pp)C_{N;\o}(\qq+\pp,\qq)
\hg^{(N)}_\o(\qq+\pp)\hac_{j}(\qq)
\;.\lb{mjmj}
\eea
We remark that  $\hU_{\o}^{(i,j)}(\pp,\qq)=\hU_{\o}^{(j,i)}(\qq,\pp)$;
in particular
$\hU_{\o}^{(i,j)}\=0$ if 
neither $j$ nor $i$ equals $N$.
As proved in \cite{[BM2]} (see also appendix \pref{A}) it is possible to decompose
\bea\lb{91}
\hU_\o^{(i,j)}(\qq+\pp,\qq)
=\sum_{\bar\o}D_{\bar\o}(\pp)
\hS_{\bar\o,\o}^{(i,j)}(\qq+\pp,\qq)
\cr\cr
\hQ_\o^{(N,j)}(\qq+\pp,\qq)
=\sum_{\bar\o}D_{\bar\o}(\pp)
\hP_{\bar\o,\o}^{(N,i)}(\qq+\pp,\qq)
\eea
for $\hS_{\bar\o,\o}^{(i,j)}$ such that, calling 
\be
S_{\bar\o,\o}^{(i,j)}(\zz;\xx,\yy)
=\int\!{d\pp\;d\qq\over (2\p)^4}\;
e^{-i\pp(\xx-\zz)}e^{-i\qq(\yy-\zz)}\hS_{\bar\o,\o}^{(i,j)}(\pp,\qq)
\ee
and similarly for $P_{\bar\o,\o}^{(N,j)}$, 
 for any positive integers $p,q$ there exists a constant
$C_{p,q}>1$ such that
\bea\lb{61}
|S_{\bar\o,\o}^{(N,j)}(\zz;\xx,\yy)|
\le C_{p,q} {\g^{N}\over 1+[\g^N|\xx-\zz|]^p}
{\g^{j}\over 1+[\g^j|\yy-\zz|]^q}
\cr\cr
|P_{\bar\o,\o}^{(N,j)}(\zz;\xx,\yy)|
\le C_{p,q} {\g^{N}\over 1+[\g^N|\xx-\zz|]^p}
{\g^{2j}\over 1+[\g^j|\yy-\zz|]^q}
\eea

The lemma holds if we choose $\n_N$ to be
\be\lb{61b}
\n_N\defi\sum_{i,j=-\io}^N
\int\!{d\pp\over (2\p)^2}\
\hS_{-\o,\o}^{(i,j)}(\pp,\pp)
\ee
As proved in \cite{[BFM]}, in the limit of removed cutoff of
\pref{61b}  equals $\l\over 2\p$. Therefore we have to prove that with
this choice \pref{89} holds true.

The integration of $\WW_\D(\a,\f)$ is done
by a multiscale integration similar to the previous one. 
After the integration of the fields $\ps^{(N)},\ldots,\ps^{(k+1)}$ 
we get the effective potential $\VV^{(k)}_\D$ such that 
\be
 e^{-\VV^{(k)}_\D(\psi^{(\le k)},\a,\f)}=
\int\! P(d\psi^{[k+1,N]})\; 
e^{-\VV^{(N)}_\D(\psi^{(\le N)},\a,\f)}
\ee
In particular, in view of \pref{der}, we are interested in the part of
$\VV^{(k)}_\D(\psi^{(\le k)},\a,\f)$ linear in $\a$,
 that we call $\KK^{(k)}_\D(\psi^{(\le k)},\a,\f)$. We first consider
 the kernels for $\f=0$.
\bea\lb{kerb}
\KK^{(k)}_\D(\ps,\a,0)
\cr\cr
=\sum_{m\ge 1} 
\sum_{\uo,\us}
\int\! d\ux d\uy d\uz\;
{K^{(1;2m)(k)}_{\D;\o,\s,\us}(\ux;\uy,\uz)\over 2m!}
\a_{\xx,\o,\s}
\prod_{i=1}^{m}\psi^{+}_{\yy,\o,\s_i}\prod_{i=1}^{m}\psi^{-}_{\zz,\o,\s_i}
\eea
As consequence of \pref{91}, we decompose 
\bea\lb{100}
\hK^{(1;2m)(k)}_{\D;\o;\s,\ss}(\pp;\uk)
\defi \sum_{\bar\o}D_{\bar\o}(\pp) 
\hW^{(1;2m)(k)}_{\D;\bar\o,\o;\s,\ss}(\pp;\uk)
\eea
We prove the following result.
\begin{lemma}\lb{l5}
For $|\l|$ small enough and $\pp,\kk,\pp+\kk\not=0$, we have that
$W^{(1;2m)(k)}_{\D;\o,\s,\us}$
analytic in $\l$ and, for $m\ge 1$,
\be\lb{kip}
\|W^{(1;2m)(k)}_{\D;\o,\s,\us}\|_k\le C|\l|\g^{-{1\over 2}(N-k)}
\g^{k(1-m)}
\ee
\end{lemma}
{\bf\0Proof.}
We integrate as in \pref{gf11}, and the difference with respect to 
\pref{gf1} is that the term $\int\!d\xx\; J_{\xx,\o,\s} \psi^+_{\xx,\o,\s}\psi^-_{\xx,\o,\s}$
is replaced by $T_{0}(\ps,\a)-\n_N T_{-}(\ps,\a)$.
%
%
%
%
The integration is done exactly as in \S 2;
we define for $0\le k\le N$
\be
\LL \hW^{(1;2)(k)}_{\D;\e\o,\o,\s,\us}(\pp;\kk)=
\hW^{(1;2)(k)}_{\D;\e\o,\o,\s,\us}(\pp;\kk)
\defi\hat\n^\e_{k,\o,\s}(\pp;\kk)
\ee
so that for $k\ge 0$
\bea\lb{h12}
\LL\VV_\D^{(k)}(\psi^{(\le k)},\a,0)=\LL\VV(\psi^{(\le k)},0)
\cr\cr
+\sum_{\e=\pm} \int {d\kk d\pp\over(2\pi)^4} 
D_{\e\o}(\pp)\hat\n^\e_{k,\o,\s}(\pp;\kk) 
\ha_{\pp,\o,\s} \hp^{(\le k)+}_{\kk,\o,\e\s}\hp^{(\le k)-}_{\kk+\pp,\o,\e\s}
\eea
$\LL\VV(\psi^{(\le k)},0)$ is given by the first two addenda of \pref{ef}.
On the other hand for $k\le 0$ we define
\be\lb{71bis}
\LL \hW^{(1;2)(k)}_{\D;\e\o,\o,\s,\us}(\pp;\kk)=
\hW^{(1;2m)(k)}_{\D;\e\o,\o,\s,\us}(0;0)
\defi \hat\n^\e_{k,\o,\s}
\ee
so that we define $\hat\n^+_{k,\o,\s}$ and
$\hat\n^-_{k,\o,\s}$ such that 
\bea\label{h11}
\LL\VV_\D^{(k)}=\LL\VV(\psi^{(\le k)},0)
\cr\cr+
\sum_{\e=\pm} \hat\n^\e_{k,\o,\s}
\int {d\kk d\pp\over(2\pi)^4} 
D_{\e\o}(\pp)
\a_{\pp,\o,\s} \psi^{(\le k)+}_{\kk,\o,\e\s}\psi^{(\le k)-}_{\kk+\pp,\o,\e\s}
\eea
Proceeding as in \S 2 we can write
\be
\hW^{(k)}_\D(\pp;\uk)
=\sum_{n=0}^\io
\sum_{\t\in\TT_{k,n}^{2,1}} \sum_{\bP\in \PP
\atop |P_{v_0}|=2} \hW^{(k)}_{\D,\t}(\pp;\uk)\;, 
\ee
where $\TT_{j,n}^{2,1}$ is a family of trees, defined as in \S 2 with the only
difference that to the end-points $v$ is now associated 
\pref{h12} for $h_v\ge 0$ or
\pref{h11} for $h_v< 0$. 

Assume that 
\bea\lb{hb2}
|\n_k^\e|\le C |\l| \g^{-{1\over2}(N-k)}\;,
\eea
then
\bea\lb{rrrt}
\|W_{\D,\t}^{(k)}\|_k\le (c\bar\e_h)^{n} \g^{-{1\over 2}(N-k)}
\g^{h(2-{|P_{v_0}|\over 2}-n^\a_{v_0})}
\cdot\cr\cr\cdot
\prod_{v\ {\rm not}\ {\rm e. p.}}\g^{-({|P_v|\over 2}-2+z_v+n^\a_v)}
\eea
where $n^\a_v$ is the number of endpoints of type $\a$ following the
vertex $v$ and, by construction, $n^\a_{v_0}=1$.  ${|P_v|\over 2}-2+z_v+n^\a_v>0$;
this formula implies immediately \pref{kip}. 

The bound \pref{61} says that, for obtaining the dimensional
bound \pref{rrrt},
the function $S^{(i,j)}$ is exactly equivalent to the contraction of
the   operator $J\ps^+\ps^-$, with one
$\psi$ field contracted on scale $i$, and the other contracted on
scale $j$. This is coherent with thinking to the external field
$D_\o(\pp)\ha_{\pp,\o,\s}$ as bearing the same dimension of the $J$
field.

To avoid the $(n!)^2$ bounds for
the truncated expectation require more care. Indeed, in the
contraction of the operator $J\ps^+\ps^-$ one propagator belongs  to
the anchored tree of formula \pref{te2}, while the other may belong 
to the anchored tree, or be inside the Gram determinant. When studying
the the contraction of the kernel $T_0$ it is convenient to avoid the
bound of the Gram determinant with \pref{17b} directly. The determinant
can be expanded with respect to the entries of one the row and the
corresponding minors; in particular, 
we choose the row (made of $l$ entries) containing the propagator
coming out of the operator $T_0$, so that, together with the other
propagator in the anchored tree,  we can reconstruct the
function $S^{(i,j)}$ times a  monomial in the parameters $\tt$ that
can be always bounded by 1; the corresponding minors are Gram determinants
of dimension $l-1$, that can be bounded as in \pref{17}. Therefore, the expansion
with respect to a row make us loose a factor $l$ with respect to the
usual bound, namely a $C^n$ factor more in the final bound.   

In order to prove \pref{hb2} we note that
\bea\lb{71ter}
W^{(1;2)(k)}_{\D;\bar\o,\o,\s,\s'}(\zz;\xx,\yy)
\cr\cr
=\sum_{i,j=k}^N\int\! d\uu d\ww\;
S^{(i,j)}_{\bar\o,\o}(\zz;\uu,\ww)W^{(0;4)(k)}_{\o;\s,\s'}(\uu,\ww,\xx,\yy)
\cr\cr
-\d_{\bar\o,-\o}\n_N\int\! d\ww\;
 v(\zz-\ww)W^{(1;2)(k)}_{\o;-\s,\s'}(\ww;\xx,\yy)
\eea
%
%
The reason for which in the second line of \pref{71ter} there is $W^{(0;4)(k)}_{\o;\s,\s'}(\uu,\ww,\xx,\yy)$
and not also  non-connected graphs whit four external legs is the following:
\bd
\item{a)} 
Defining $(1-\c_N(\kk))f_i(\kk)=\d_{i,N}u_N(\kk)$,
the graphs in which one between the fields $\hp$ in $T_0$
is contracted with a kernel $\hW_\s^{(0;2)(k)}$ is of the form:
\bea\lb{67c}
{\c_N(\kk+\pp)-1\over \c_N(\kk+\pp)}
D_\o(\kk+\pp)\hg_\o(\kk)\hW_\s^{(0;2)(k)}(\kk)
\cr\cr-
u_N(\kk)\hW^{(0;2)(k)}_\s(\kk)
\eea
This term is not compatible with the structure of the
multiscale expansion of the Schwinger functions, since by support
properties we have $|\kk+\pp|,|\kk|> \g^N$ while, by construction,
the fields $\ps^{\le N-1}_{\kk+\pp}$ and $\ps^{\le N-1}_{\kk}$, 
implies the constraint $|\kk+\pp|,|\kk|<\g^N$.
\item{b)}  The graphs in which all and two the fields $\hp$
in $T_0$ are contracted, each one with its own $\hW^{(0;2)(k)}_\s$ have the form
\bea\lb{67d}
-\lft[u_N(\kk+\pp)\hg_\o(\kk)-u_N(\kk)\hg_\o(\kk+\pp)\rgt]
\cdot\cr\cr\cdot
\hW_\o^{(0;2)(k)}(\kk)\hW_\s^{(0;2)(k)}(\kk+\pp)
\eea
hence they are not compatible with the multiscale expansion for the
very same reason as above.
\ed

\pref{71ter} is analyzed in a way similar to the one followed in \S 3;
by using the decomposition of $W^{(0;4)(k)}_{\o,\s}$ in Fig. \ref{q7},
so obtaining the decomposition for $W^{(1;2)(k)}_{\D;\bar\o,\o,\s,\s'}$
depicted  in Fig.\ref{q8b}.
\insertplot{320}{230}
{\ins{107pt}{204pt}{$\zz$}
\ins{145pt}{194pt}{$\uu$}
\ins{145pt}{232pt}{$\ww$}
\ins{170pt}{196pt}{$\xx$}
\ins{170pt}{231pt}{$\yy$}

\ins{210pt}{216pt}{$-\n_N$}

\ins{237pt}{204pt}{$\zz$}
\ins{255pt}{204pt}{$\ww$}
\ins{285pt}{196pt}{$\xx$}
\ins{285pt}{231pt}{$\yy$}

\ins{40pt}{159pt}{$=$}
\ins{67pt}{146pt}{$\zz$}
\ins{117pt}{146pt}{$\uu$}
\ins{137pt}{146pt}{$\ww$}
\ins{165pt}{140pt}{$\xx$}
\ins{165pt}{177pt}{$\yy$}
\ins{70pt}{180pt}{(a)}

\ins{210pt}{159pt}{$-\n_N$}
\ins{227pt}{146pt}{$\zz=\uu$}
\ins{257pt}{146pt}{$\ww$}
\ins{285pt}{140pt}{$\xx$}
\ins{285pt}{177pt}{$\yy$}
\ins{240pt}{180pt}{(b)}

\ins{7pt}{99pt}{$+$}
\ins{32pt}{88pt}{$\zz$}
\ins{62pt}{124pt}{$\uu$}
\ins{108pt}{122pt}{$\uu'$}
\ins{108pt}{78pt}{$\ww$}
\ins{87pt}{90pt}{$\ww'$}
\ins{120pt}{115pt}{$\yy$}
\ins{120pt}{82pt}{$\xx$}
\ins{30pt}{120pt}{(c)}

\ins{157pt}{99pt}{$-$}
\ins{182pt}{88pt}{$\zz$}
\ins{210pt}{76pt}{$\ww$}
\ins{243pt}{79pt}{$\ww'$}
\ins{252pt}{88pt}{$\uu$}
\ins{262pt}{91pt}{$\uu'$}
\ins{285pt}{82pt}{$\xx$}
\ins{285pt}{115pt}{$\yy$}
\ins{180pt}{120pt}{(d)}

\ins{-10pt}{29pt}{$+\d_{\s,\s'}$}
\ins{32pt}{18pt}{$\zz$}
\ins{72pt}{6pt}{$\ww$}
\ins{102pt}{6pt}{$\xx$}
\ins{97pt}{35pt}{$\ww'$}
\ins{78pt}{55pt}{$\uu=\yy$}
\ins{30pt}{50pt}{(e)}

\ins{140pt}{29pt}{$-\d_{\s,\s'}$}
\ins{182pt}{18pt}{$\zz$}
\ins{222pt}{6pt}{$\ww$}
\ins{252pt}{6pt}{$\xx$}
\ins{247pt}{35pt}{$\ww'$}
\ins{237pt}{55pt}{$\uu$}
\ins{254pt}{58pt}{$\uu'$}
\ins{286pt}{55pt}{$\yy$}
\ins{180pt}{50pt}{(f)}
}
{q8b}{\lb{q8b}: Graphical representation of  \pref{71bis}}{0}
\\
Fixed the integer $q$ and 
calling $b_j(\xx)\defi C_q\g^j/(1+[\g^j|\xx|]^q)$,
we bound the r.h.s. member 
in the same spirit as in \S 3.
\bd
\item{1.}  Graphs (c) and (d) are:
\bea
\sum_{i,j=k}^N\int\! d\uu d \uu' d\ww d\ww'\;
S^{(i,j)}_{-\o,\o}(\zz;\uu,\ww)
g_\o(\uu-\uu')v(\uu-\ww')
\cdot\cr\cr\cdot
W^{(1;4)(k)}_{\o;-\s,\s,\s'}(\ww';\uu',\ww,\xx,\yy)
\cr\cr
+\sum_{i,j=k}^N\int\! d\uu d \uu' d\ww d\ww'\;
S^{(i,j)}_{-\o,\o}(\zz;\uu,\ww)
W^{(0;2)(k)}_{\o;\s}(\ww,\ww') 
\cr\cr\cdot
g_\o(\ww'-\uu)v(\uu-\uu')W^{(1;2)(k)}_{\o;-\s,\s'}(\uu';\xx,\yy)
\eea
Since either $i$ or $j$ has to be $N$, and 
by the bound \pref{61},
the norm of (c) is bounded by 
\bea
2|v|_\io\sum_{j,m=k}^N\int\!  d \xx  d \uu' d\ww d\ww'\;
|W^{(1;4)(k)}_{\o;-\s,\s,\s'}(\ww';\uu',\ww,\xx,\yy)|
\cr\cdot
\int\!d\zz d\uu \;
b_N(\zz-\uu) b_j(\zz-\ww) |g^{(m)}_\o(\uu-\uu')|
\eea
and hence we can clearly proceed as 
for \pref{27} but now
the scale of higher momenta 
is fixed to be $N$, and therefore we get the bound
\bea
C_1 |\l|\cdot |v|_\io\cdot  \g^{-k} \sum_{i=k}^N \sum_{i'=k}^i
\g^{-N}\g^{-i}\g^{i'}
\cr\cr\le
C_2 |\l|\g^{-k-N}(N-k)
\le C_3
|\l| \g^{-2k}\g^{-(1/2)(N-k)} 
\eea
A similar bound can be obtained for (d).
\item{2.} The graphs (e) and (f) are:
\bea
\d_{\s,\s'}\sum_{i,j=k}^N\int\! d\uu  d\ww d\ww'\;
S^{(i,j)}_{-\o,\o}(\zz;\uu,\ww)
W^{(1;2)(k)}_{\o;\s,\s}(\ww';\ww,\xx)
\cr\cr\cdot
\lft[ \d(\uu-\yy)+
\int\! d \uu'\;g_\o(\uu-\uu')v(\yy-\ww')
W^{(0;2)(k)}_{\o;\s,\s}(\uu',\yy)\rgt]
\eea
The bound for the 
graph (e) is $C|v|_{\io}\cdot\|W^{(1;2)(k)}_{\o;\s,\s}\|_k
\cdot|b_N|_1\sum_{j=k}^N|b_j|_1\le C \g^{-(N-k)}\g^{-2k}$.  
Similar bound holds for (f).
\item{3.} The graphs (a) and (b) are:
\bea\lb{75}
\int\! d\uu\;
\lft[\sum_{i,j=k}^N
S^{(i,j)}_{-\o,\o}(\zz;\uu,\uu)
-\n_N \d(\zz-\uu)\rgt]
\cdot\cr\cr\cdot
\int\!d\ww\; v(\uu-\ww)W^{(1;2)(k)}_{\o;-\s,\s'}(\ww,\xx,\yy)
\eea
Using the identity \pref{idb}, 
for graph (a) we have
\bea\lb{125b}
\sum_{i,j=k}^N\int\! d\uu d\ww \;
S^{(i,j)}_{-\o,\o}(\zz;\uu,\uu)v(\uu-\ww)
W^{(1;2),(k)}_{\o;-\s,\s'}(\ww;\xx,\yy)
\cr\cr=
\int\! d\ww\;  v(\zz-\ww)
W^{(1;2),(k)}_{\o;-\s,\s'}(\ww;\xx,\yy)\sum_{i,j=k}^N\int\! d\uu \;
S^{(i,j)}_{-\o,\o}(\zz;\uu,\uu)
\cr\cr
+
\sum_{p=0,1}\sum_{i,j=k}^N\int\! d\uu \;
S^{(i,j)}_{-\o,\o}(\zz;\uu,\uu)(u_p-z_p)
\cdot\cr\cr\cdot
\int_0^1\!d\t\;
\int\! d\ww\; (\dpr_pv)(\zz-\ww+\t(\uu-\zz))
W^{(1;2),(k)}_{\o';-\s,\s'}(\ww;\xx,\yy)
\eea
The latter term is irrelevant and
vanishing in the limit $N-k\to+\io$: using that one 
between $i$ and $j$ is on scale $N$,
a bound for its norm is
\be
2\|W^{(1;2),(k)}_{\o';-\s,\s'}\|_k\cdot
|\dpr v|_1\cdot
\sum_{j=k}^N\int\! d\uu \;
b_N(\zz-\uu) b_j(\zz-\uu)|(u_p-z_p)|
\ee
and 
we obtain the bound $C \g^{-k}\g^{-(N-k)}$.
The former term in the r.h.s. member of \pref{125b}
is compensated by (b).
Indeed we have
\bea\lb{78bis}
\sum_{i,j=k}^N\int\! d\uu \;
S^{(i,j)}_{-\o,\o}(\zz;\uu,\uu)-\n_N=
2\sum_{j\le k-1}\int\! d\uu \;
S^{(N,j)}_{-\o,\o}(\zz;\uu,\uu)
\eea
and hence the bound for such a difference is $C\g^{-(N-k)}$.
\ed 
The graph expansion for
$W^{(1;2)(k)}_{\D;\o,\o,\s,\s'}$ is again given by Fig.\ref{q8b},
but for $\n_N$ replaced by $0$. Hence a bound can be obtained 
with the same above argument, with only one important difference: 
the contribution that in the previous analysis were compensated 
by (b) now are zero by symmetries. Indeed,  
calling $\kk^*$ the rotation of $\kk$ of $\p/2$ and 
since $\hS^{(i,j)}_{\bar\o,\o}(\kk^*,\pp^*)=-\o\bar\o\hS^{(i,j)}_{\bar\o,\o}(\kk,\pp)$,
in place of the bound \pref{78bis}, in this case we have:
\be\lb{79}
\sum_{i,j=k}^N\int\! d\uu \;
S^{(i,j)}_{\o,\o}(\zz;\uu,\uu)=
\sum_{i,j=k}^N\int\! {d\kk\over (2\p)^2} \;
\hS^{(i,j)}_{\o,\o}(\kk,-\kk)=0
\ee
Finally, so far we have obtained \pref{hb2} for $k\ge 0$.

Let us consider, now,  the case $k<0$.
By \pref{71bis} we have
\bea\lb{84}
\LL\hW^{(1;2)(k)}_{\D;\bar\o,\o,\s,\s'}(\pp;\kk)
=\hW^{(1;2)(k)}_{\D;\bar\o,\o,\s,\s'}(0;0)
\cr\cr=
\int\!{d\qq\over (2\p)^2}\;
\hS_{\bar\o,\o}^{(i,j)}(\qq,\qq)\hW^{(0;4)(k)}_{\o;\s,\s'}(\qq,\qq,0)
-\d_{\bar\o,-\o}\n_N\; \hW^{(1;2)(k)}_{\o;-\s,\s'}(0,0)
\eea
As we noticed in \S 2
$\hW^{(1;2)(k)}_{\o;-\s,\s'}(0,0)=\d_{-\s,\s'}$;
furthermore, under a rotation of $\p/2p$,
\bea
\hS_{\bar\o,\o}^{(i,j)}(\pp^*,\qq^*)=
e^{-i(\o+\bar\o){\p\over 2p}}\hS_{\bar\o,\o}^{(i,j)}(\pp,\qq)
\cr\cr
\hW^{(0;4)(k)}_{p,\uo}(\pp^*,\qq^*,0)
=
e^{-i\o\p(1-{1\over p})}\hW^{(0;4)(k)}_{p,\uo}(\pp,\qq,0)
\eea
hence the integral in \pref{84} is non-zero 
only for $p=1$ and $\bar\o=-\o$, case in which 
\pref{84} is reduces to \pref{78bis}. 
\qed
\vskip.5cm
We can finally discuss the bound for $R^{(1;2)}_{\o;\s',\s}(\pp;\kk)$
so finally proving Lemma \ref{l5}.
It can be
written by a sum of trees essentially identical to the ones for
$\hG^{(1;2)}_{\o;\s',\s}(\pp;\kk)$, with the only important difference that
there are three different special endpoints associated to the
field $\a$, corresponding to the three different terms in
\pref{vvn}; we call these endpoints of type $T_+,T_-,T_0$
respectively.

The sum over the trees such that the endpoint is of type $\n^\pm_{k,\o,\s}$
can be bounded as in \pref{2.61}, the only difference being that,
thanks to the bound \pref{hb2},
one has to multiply the r.h.s. by
a factor $|\l|\g^{-{1\over 2}(N-k)}$, for $k$ the scale of the
endpoint. This factor has to be inserted also in
the r.h.s. of the bounds \pref{2.62},  hence, it is
easy to see that the contributions of these trees vanishes as
$N\to\io$.

Let us now consider the trees with an endpoint of type $T_0$. The
fields of the $T_0$ endpoint are contracted at scale $j,N$; this
implies that  $h_J=N$: since
$d_v+r_v-1/4>0$ for all vertices belonging to the path connecting the
endpoint to the root, we can replace in  the r.h.s. of the bounds
\pref{2.62} $d_v+r_v$  with $d_v+r_v-1/4$ and add a factor 
$\g^{-(N-h_\kk)/4}$, so that
\be
\lim_{N\to\io}R^{(1;2)}_{\o;\s',\s}(\pp;\kk)=0
\ee
\*
\section{The Closed Equation}
By \pref{mas1b}, for $k=-\io$, we obtain
the  Schwinger-Dyson equation for the two-point Schwinger function
\be\lb{sde}
\hS_{N;\o,\s}(\kk)=\hat g_\o(\kk)
\lft[1-\l  
\int\!{d\pp\over (2\p)^2}\ 
\hat v(\pp) \hG_{N;\o,-\s;\s}(\pp;\kk)\rgt]
\ee
We define
$$
a_N(\pp)={1\over D_{\o}(\pp)-\n_N \hv(\pp)D_{-\o}(\pp)}\quad\quad
\bar a_N(\pp)={1\over D_{\o}(\pp)+\n_N \hv(\pp)D_{-\o}(\pp)}
$$
and summing over $\s'$  the equation,  we obtain the 
{\it vector Ward Identity} (associated the phase symmetry):
\bea
\sum_{\s'} \hG_{N;\o,\s';\s}(\pp;\kk)
=a_N(\pp)\sum_{\e,\bar\o} D_{\bar\o}(\pp)\hR^{(1;2)}_{\bar\o,\o;\e\s,\s}(\pp;\kk)
\cr\cr+
a_N(\pp)
\Big[\hS_{N;\o,\s}(\kk)-\hS_{N;\o,\s}(\kk+\pp)\Big]
\eea
while multiplied times $\s'$ the equation, and summing over $\s'$,
we obtain the {\it axial Ward Identity} (associated
to the chiral symmetry):
\bea
\sum_{\s'}\s' \hG_{N;\o,\s';\s}(\pp;\kk)=
\s\bar a_N(\pp)\sum_{\e,\bar\o}\e
D_{\bar\o}(\pp)\hR^{(1;2)}_{\bar\o,\o;\e\s,\s}(\pp;\kk)
\cr\cr+
\s\bar a_N(\pp)\Big[\hS_{N;\o,\s}(\kk)-\hS_{N;\o,\s}(\kk+\pp)\Big]
\eea
Finally, from these two equations, since ${1+\r\s'\over 2}=\d_{\r,\s'}$
\bea\lb{wi}
\hG_{N;\o,\s';\s}(\pp;\kk)
=\sum_{\e,\bar\o} {a_N(\pp)+\bar a_N(\pp)\e\over 2}
D_{\bar\o}(\pp)\hR^{(1;2)}_{\bar\o,\o;\e\s',\s}(\pp;\kk)
\cr\cr+
{a_N(\pp)+\s\s'\bar a_N(\pp)\over 2}
\Big[\hS_{N;\o,\s}(\kk)-\hS_{N;\o,\s}(\kk+\pp)\Big]
\eea
In order to shorten the notation we now define 
\be\lb{Ae}
\hA_\e(\pp)\defi 
{\hv(\pp)[a(\pp)+\e\bar a(\pp)]\over 2}\;.
\ee
Let 
$$
\hR^{(2)}_{\o;\e;\s}(\kk)
\defi
\sum_{\bar\o}\int\!{d\pp\over (2\p)^2}\ 
\bar\c_N(\pp)\hA_-(\pp)
D_{\bar\o}(\pp)\hR^{(1;2)}_{\bar\o,\o;-\e\s;\s}(\pp;\kk)
$$
\begin{theorem}\lb{t42}
If  $|\l|$ is small enough and for fixed momentum $\kk$, in the limit $N\to \io$
we obtain
\bea\lb{ce}
D_\o(\kk)\hS_{\o,\s}(\kk)=
1+\l
\int\!{d\pp\over (2\p)^2}\ 
\hA_-(\pp)
\hS_{\o,\s}(\kk+\pp)
\eea
\end{theorem}
By solving \pref{ce} (see appendix B) and using \pref{87},
Theorem \ref{th3} follows.
In order to 
prove of Theorem \ref{t42} we have to show that
\bea\lb{131}
\lim_{N\to\io}\hR^{(2)}_{\o;\e;\s}(\kk)=0
\eea
it is convenient to write
\be\lb{78}
\hR^{(0;2)}_{\o;\e;\s}(\kk)
={\partial^2 
\WW_{T,\e} \over 
\dpr \hb_{\kk,\o,\s}\dpr\hf^-_{\kk,\o,\s}}(0)
\ee
where we have introduced the new generating functional
\bea
e^{\WW_{T,\e} (\b,\f)} 
=\int\! P(d\psi^{(\le N) })\;
e^{-\VV^{(N)}_{T,\e}(\ps^{\le N},\b,\f)}
\cr\cr
\defi\int\! P(d\ps)\;
\exp\lft\{ 
-\l V(\ps^{(\le N) })+\left[T^{(\e)}_{1} -\n_{N} 
T^{(\e)}_{-}\right]\left(\ps^{(\le N) }, \b\right)\rgt\}
\cr\cr
\exp\lft\{ 
\sum_{\o,\s}\int\!d\xx\;
[\f^{+}_{\xx,\o,\s}\ps^{(\le N)- }_{\xx,\o,\s}
+\ps^{(\le N)+ }_{\xx,\o,\s}\f^{-}_{\xx,\o,\s}]\rgt\}
\eea
with
\bea\lb{80}
&&
T^{(\e)}_{1}(\psi,\b)=\sum_{\o,\s}
\int\!{d\pp\;d\qq\over (2\p)^4}\;
\bar\c_N(\pp)\hA_{\e}(\pp)C_{N;\o}(\qq+\pp,\qq)
\cdot\cr\cr
&&\phantom{*************}\cdot
\hb_{\kk,\o,\s}\hp^-_{\kk+\pp,\o,\s}
\hp^+_{\qq+\pp,\o,-\e\s}\hp^-_{\qq,\o,-\e\s}
\cr\cr
&&
T^{(\e)}_{-}(\psi,\b)=\sum_{\o,\s}
\int\!{d\pp\;d\qq\over (2\p)^4}\; 
\bar\c_N(\pp)\hA_{\e}(\pp)\hv(\pp)D_{-\o}(\pp)
\cdot\cr\cr
&&\phantom{*************}\cdot
\hb_{\kk,\o,\s}\hp^-_{\kk+\pp,\o,\s}
\hp^+_{\qq+\pp,\o,\e\s}\hp^-_{\qq,\o,\e\s}
\eea
and $\n_N$ is defined in the previous section.

The integration of 
$\WW_{T,\e}$ can be done in a way very similar to the previous ones.
After the integration of the fields $\ps^{(N)},\ldots, \ps^{(k+1)}$,
we get
\be
e^{-\VV^{(k)}_{T,\e}(\ps^{(\le k)},\b,\f)}\defi \int\!
P(d\psi^{[k+1,N] })\;
e^{-\VV^{(N)}_{T,\e}(\ps^{(\le N)},\b,\f)}
\ee
and we call $\HH_{T,\e}^{(k)}$ the part of 
$\VV^{(k)}_{T,\e}$ that is linear in $\b$
\bea
\HH^{(k)}_{T,\e}(\ps,0,\b)\lb{ker3}
\cr\cr
=\sum_{m\ge 1} 
\sum_{\uo,\us}
\int\!d\zz d\xx d\yy \; 
{H^{(1;2m+1)(k)}_{T,\e;\o,\us}(\zz;\xx,\yy)\over 2m!}
\b_{\zz,\o,\s'}
\prod_{i=1}^{m}\ps^{+}_{\xx_i,\o,\s_i}
\prod_{i=1}^{m+1}\ps^{-}_{\yy_i,\o,\s_i}
\eea
\begin{theorem}\lb{lb3}
If $|\l|$ small enough,  for any $h:k+1\le h\le N$,
\bea
\lb{pcb}
 \|H^{(1;2m+1)(k)}_{T,\e;\us,\uo}\|_k
\le C \g^{-{1\over 2}(N-k)} \g^{k(1-m)}
\eea
\end{theorem}
{\bf \0Proof.}
The integration is done exactly as in \S 2;
we define for $0\le k\le N$
\bea
\LL \hH^{(1;1)(k)}_{T,\e;\us,\uo}(\pp;\kk)=
\hH^{(1;1)(k)}_{T,\e;\us,\uo}(\pp;\kk)\defi \hat z^\e_k(\pp;\kk)
\cr\cr
\LL \hH^{(1;3)(k)}_{T,\e;\us,\uo}(\pp;\uk)=
\hH^{(1;3)(k)}_{T,\e;\us,\uo}(\pp;\uk)\defi \hat \l^\e_k(\pp;\uk)
\eea
so that for $k\ge 0$ 
\bea\lb{h12b}
\LL\VV_{\TT,\e}^{(k)}=\LL\VV(\psi^{(\le k)},0)
\cr\cr
+ \int {d\kk d\pp d\qq \over(2\pi)^4} 
\hat\l^\e_k(\kk,\pp,\qq) 
\hb_{\pp,\o,\s} \hp^{(\le k)-}_{\kk+\pp,\o,\s}
\hp^{(\le k)+}_{\kk,\o,-\s}\hp^{(\le k)-}_{\kk+\pp,\o,-\s}
\cr\cr
+ \int {d\kk \over(2\pi)^2} 
\hat z^\e_k(\kk) 
\hb_{\kk,\o,\s} \hp^{(\le k)-}_{\kk,\o,\s}
\eea
where $\LL\VV(\psi^{(\le k)},0)$ is given by the first two addenda of \pref{ef}.

On the other hand for $k\le 0$ we define
\bea
\LL \hH^{(1;1)(k)}_{T,\e;\us,\uo}(\pp;\kk)=
\hH^{(1;1)(k)}_{T,\e;\us,\uo}(0;0)\equiv \tilde z^\e_k
\cr\cr
\LL \hH^{(1;3)(k)}_{T,\e;\us,\uo}(\pp;\uk)=
\hH^{(1;3)(k)}_{T,\e;\us,\uo}(0;0)=\tilde \l^\e_k
\eea
so that for $h<0$
\bea\lb{h11b}
\LL\VV_{\TT,\e}^{(k)}=\LL\VV(\psi^{(\le k)},0)
\cr\cr
+ \hat\l^\e_k \int {d\kk d\pp d\qq \over(2\pi)^4} 
\hb_{\pp,\o,\s} \hp^{(\le k)-}_{\kk+\pp,\o,\s}
\hp^{(\le k)+}_{\kk,\o,-\s}\hp^{(\le k)-}_{\kk+\pp,\o,-\s}
\cr\cr
+  \hat z^\e_k \int {d\kk \over(2\pi)^2}
\hb_{\kk,\o,\s} \hp^{(\le k)-}_{\kk,\o,\s}
\eea
where $\LL\VV(\psi^{(\le k)},0)$ is given by the first two addenda of \pref{loc}.
Proceeding as in \S 2 we can write
\be
H^{(1;2m+1)(k)}_{T,\e;\us,\uo}=
\sum_{n=0}^\io\sum_{\t\in\TT_{k,n}}\sum_{\bf P} H^{(1;2m+1)(k)}_{T,\e;\t,{\bf P}}
\ee
where $\TT_{k,n}$ is a family of trees, defined as in \S 2 with the only
difference that to the end-points $v$ is now associated 
\pref{h12b} for $h_v\ge 0$ or
\pref{h11b} for $h_v< 0$; and there is one  special endpoint with
field $\b$.

Assume that, for any $k$,
\bea\lb{147}
||\l^\e_k||_k,|| z^\e_k||_k\le C |\l| \g^{-{1\over2}(N-k)}\;,
\eea
then, proceeding as above
\bea\lb{rrrq}
\|H_{T,\e;\t,\bP}^{(k)}\|_k\le (c\bar\e_{k+1})^{n-n^\a_{v_0}} \g^{-{1\over2}(N-k)}
\g^{h(2-{|P_{v_0}|\over 2}-n^\a_{v_0})}
\cdot\cr\cr\cdot
\prod_{v\ {\rm not}\ {\rm e. p.}}\g^{-({|P_v|\over 2}-2+z_v+n^\a_v)}
\eea
and again ${|P_v|\over 2}-2+z_v+n^\a_v>0$. 
In order to prove \pref{147} we can write
\be\lb{fffg}
H^{(1;3)(k)}_{T,\e;\o,\o',\s}=H^{a(1;3)(k)}_{T,\e;\o,\o',\s}+H^{b(1;3)(k)}_{T,\e;\o,\o',\s}
\ee
where:
\bd
\item{1.}
$H^{a(1;3)(k)}_{T,\e;\o,\us}$ contains the term in which 
the field $\hp_{\kk+\pp,\o,\s}$ 
of $T_1$ and $T_-$ is not contracted or is contracted with a $\hW^{(0;2)(k)}$:
\bea
\hH^{a(1;3)(k)}_{T,\e;\o,\us}(\kk,\pp)
\cr\cr
=
\lft[1+\hg^{[k+1,N]}_\o(\pp)\hW^{(0;2)(k)}(\pp)\rgt]\hA_\e(\pp) 
D_\o(\pp)\hW^{(1;2)(k)}_{\D,\e;\o,\o',\s}(\kk;\pp)
\eea
 for
$k\ge 0$ we have already proved the 
bound $\|W^{(0;2)(k)}\|_k\le C|\l|^2\g^{-k}$;
for $k< 0$, we use the fact that the local part of
(up to first order of Taylor expansion in $\kk$) 
$\hW^{(0;2)(k)}$ is zero, and the rest has a dimensional gain of one degree;
by \pref{kip},
\be
\|H^{a(1;3)(k)}_{T,\e;\o,\us}\|_k\le C|\l|\g^{-{1\over 2}(N-k)}
\ee
For $k\le 0$ we have defined 
$\LL \hH^{a(1;3)(k)}_{T,\e;\o,\us}(\kk,\pp)=\hH^{a(1;3)(k)}_{T,\e;\o,\us}(0,0)$
and we know by symmetry that $\hH^{a(1;3)(k)}_{T,\e;\o,\us}(0,0)=0$.
\item{2.}
$H^{b(1;3)(k)}_{T,\e;\o,\us}$ contains the term in which 
the field $\hp_{\kk+\pp,\o,\s}$ 
of $T_1$ and $T_-$ is contracted. 
\insertplot{300}{120}
{
\ins{155pt}{96pt}{$-\n_N$}
\ins{151pt}{25pt}{$+$}
\ins{21pt}{25pt}{$+$}

\ins{23pt}{114pt}{$(a)$}
\ins{187pt}{114pt}{$(b)$}
\ins{39pt}{44pt}{$(c)$}
\ins{180pt}{44pt}{$(d)$}
}
{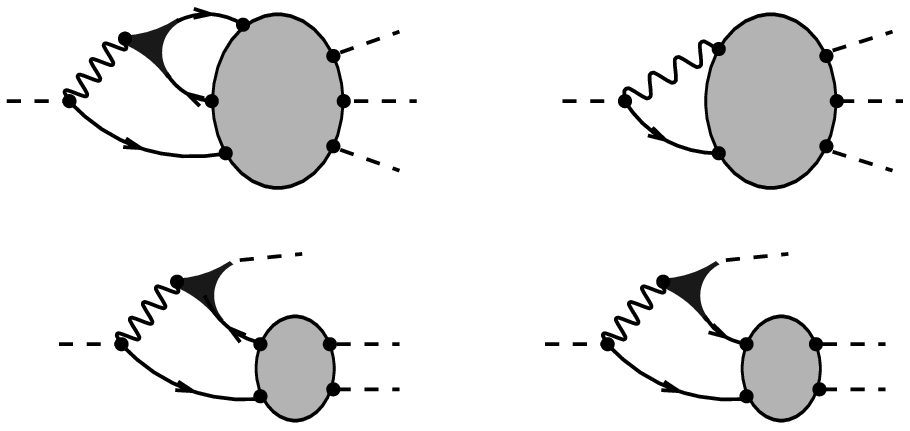}{\lb{q12}: Graphical representation of  $H^{b(1;3)(k)}_{T,\e;\o,\us}$}{0}
We can further distinguish them 
as in Fig \ref{q12}; we can write
\bea
H^{b(1;3)(k)}_{T,\e;\o,\us}(\xx,\yy, \uu,\vv)
\cr\cr=
\int d\zz d\ww\; \bar v(\xx-\zz) g_\o^{[k+1,N]}(\xx-\ww)
K^{(1;4)(k)}_{\D,\e;\o,\o',\s}(\zz;\ww,\yy, \uu,\vv)
\eea
where
$$\bar v(\xx)=\int d\pp\;
e^{i\pp\xx}\hA_\e(\pp)D_\o(\pp)$$
so that, by the bounds for $\|K^{(1;4)(k)}_{\D,\e;\o,\o',\s}\|_k$, 
$|\bar v|_\io$ and $|g^{(j)}_\o|_1$,
\be
\|H^{b(1;3)(k)}_{T,\e;\o,\us}\|_k\le C |\l|\g^{-k}\g^{{1\over 2}(N-k)}
\ee
While for $k<0$ we have that the local part of the graph is zero by
transformation under rotation. 
\ed
We consider now the terms contributing to $H^{(1;1)(k)}_{T,\e;\o,\s}$.
\bd
\item{1.} The contraction of the field $\hp^+_{\qq+\pp,\o,-\e\s}$ with 
$\hp^-_{\qq,\o,-\e\s}$ of $T_1$, possibly through a kernel 
$\hW^{(0;2)(k)}(\qq)$,
can only happen for $\pp=0$, and therefore it is forbidden 
by $\bar\c_N(\pp)$. 
\item{2.} The contraction of $\hp^+_{\qq+\pp,\o,-\e\s}$ with 
$\hp^-_{\kk+\pp,\o,\s}$ (that can take place only 
for $\e=-$), possibly through a kernel $\hW^{(0;2)(k)}(\qq+\pp)$, and 
possibly with $\hp^-_{\qq,\o,-\e\s}$ contracted with a second 
kernel $\hW^{(0;2)(k)}(\qq)$, has the following expression
\bea\lb{95}
\hH^{a(1;1)(k)}_{T,\e;\o,\s}(\kk)=\int\!{d\pp\over (2\p)^2}\;\bar \c_N(\pp+\kk) \hA_-(\pp+\kk)
\hv(\kk+\pp)u_N(\pp)
\cdot\cr\cr\cdot
\lft[1+\hg^{[k+1,N]}_\o(\pp)\hW^{(0;2)(k)}(\pp)\rgt]
\lft[1+\hg^{[k+1,N]}_\o(\kk)\hW^{(0;2)(k)}(\kk)\rgt]
\eea
For $0\le k\le N$, we define $\LL H^{a(1;1)(k)}_{T,\e;\o,\s}(\kk)
=H^{a(1;1)(k)}_{T,\e;\o,\s}(\kk)$ for such terms;
since $|\kk|$ is fixed by hypothesis, $|\pp+\kk|\le
C\g^{N}$ a bound for \pref{95} is
\bea\lb{95b}
|v|_\io\g^{-k} \g^{-(N-k)}
\lft[1+C\sum_{j=k}^N \g^{-(j-k)}\rgt]
\lft[1+C\g^{-(N-k)}\rgt]
\cr\cr\le C \g^{-k} \g^{-(N-k)}
\eea
On the other hand, for $k<0$, $\LL \hH^{a(1;1)(k)}_{T,\e;\o,\s}(\kk)
=\hH^{a(1;1)(k)}_{T,\e;\o,\s}(0)$ and 
\bea\lb{97}
\hH^{a(1;1)(k)}_{T,\e;\o,\s}(0)=
\sum_{\o'}D_{\o'}(\kk)\int\!{d\pp\over (2\p)^2}\;\bar \c_N(\pp) \hA_-(\pp)
\hv(\pp)\big(\dpr_{\o'}u_N\big)(\pp)
\cdot\cr\cr\cdot
\lft[1+\hg^{[k+1,N]}_\o(\pp)\hW^{(0;2)(k)}(\pp)\rgt]
\lft[1+\hg^{[k+1,N]}_\o(\kk)\hW^{(0;2)(k)}(\kk)\rgt]
\eea
Such an integral is zero. Indeed, we have
\bea\lb{a-}
\hA_-(\pp)={\n_N\hv^2(\pp) D_{-\o}(\pp)\over D^2_\o(\pp)-\n^2_N\hv^2(\pp)
  D^2_{-\o}(\pp)}
\cr\cr
={\n_N\hv^2(\pp) D_{-\o}(\pp)\over D^2_\o(\pp)}
\sum_{p\ge 0} \lft({\n_N\hv(\pp)
  D_{-\o}(\pp)\over D_\o(\pp)}\rgt)^{2p} \defi
\sum_{p\ge 0}\hA_{p,-}(\pp)
\eea
Under a rotation of  an angle $\th$, we have:
\bea\lb{101}
\hA_{p,-}(\pp^*)=e^{-i\o\th (4p+ 3)}
\hA_{p,-}(\pp)
\cr\cr
\hW^{(0;2)(k)}_{p',\o,\s}(\pp^*)
=
e^{-i\o\th(2p'-1)}\hW^{(0;2)(k)}_{p',\o,\s}(\pp)
\eea
and therefore, since $(4p+4+2p')>0$, 
taking $\th:\th(4p+4+2p')<2\p$, the integral \pref{97}, 
with $\hA_{-}(\pp)$ and $\hW^{(0;2)(k)}_{\o,\s}(\pp)$
replaced by $\hA_{p,-}(\pp)$ and $\hW^{(0;2)(k)}_{p',\o,\s}(\pp)$ 
respectively, is zero.
\item{3.} The contraction of $T_1$ with all and three fields
  contracted with the same kernel $\hW^{(0;4)(k)}_{\o,\us}$; and the 
contraction of $T_-$. They are:
\bea
\hH^{b(1;1)(k)}_{T,\e;\o,\s}(\kk)
\cr\cr=\sum_{\bar\o}\sum_{i,j=k}^N\int\!{d\pp\;d\qq\over (2\p)^4}\;
\bar\c_N(\pp)\hA_\e(\pp)D_{\bar\o}(\pp)\hS^{(i,j)}_{\bar\o,\o}(\qq+\pp,\qq)
\cdot\cr\cr\cdot
\hg_\o(\pp+\kk)\hW^{(0;4)(k)}_\us(\kk+\pp,\qq+\pp,\qq)
\cr\cr
+\int\!{d\pp\over (2\p)^2}\;
\bar\c_N(\pp)\hA_\e(\pp)\hv(\pp)D_{-\o}(\pp)
\cdot\cr\cr\cdot
\hg_\o(\pp+\kk)\hW^{(1;2)(k)}_\us(\kk+\pp,\pp)
\eea
\insertplot{300}{60}
{
\ins{155pt}{36pt}{$-\n_N$}
}
{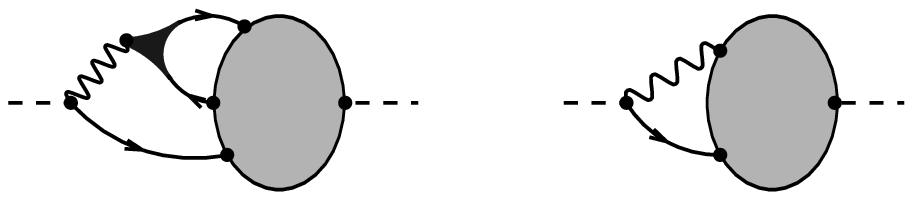}{\lb{q9b}: Graphical representation of $H^{b(1;1)(k)}_{T,\e;\o,\o',\s}$}{0}
It has a bound as \pref{hb2} times a further factor 
\be
|v|_\io\cdot \sum_{j=k}^N|g_\o^j|_1\le C\g^{-k}
\ee
For $k<0$, we have $H^{b(1;1)(k)}_{T,\e;\o,\s}(0)=0 $.
This follows 
using \pref{a-}, \pref{101} and
\bea\lb{a+}
\hA_+(\pp)={\hv(\pp)D_{\o}(\pp)\over D^2_\o(\pp)-\n^2_N\hv^2(\pp)
  D^2_{-\o}(\pp)}\defi
\sum_{p\ge 0}\hA_{p,+}(\pp)
\eea
\bea\lb{103}
\hA_{p,+}(\pp^*)=e^{-i\o\th (4p+ 1)}
\hA_{p,+}(\pp)
\cr\cr
\hW^{(0;4)(k)}_{p',\us}(\pp^*)
=
e^{-i\o\th(2p'-2)}\hW^{(0;4)(k)}_{p',\us}(\pp)
\cr\cr
D_{\bar\o}(\pp^*)\hS^{(i,j)}_{\bar\o,\o}(\qq^*+\pp^*,\qq^*)
=
e^{-i\o\th}
D_{\bar\o}(\pp)\hS^{(i,j)}_{\bar\o,\o}(\qq+\pp,\qq)
\eea
\ed
This completes the proof.
\qed

\appendix
\section{Bounds for the $\D$ function}\lb{A}
Because of the symmetry
$\hS^{(i,j)}_{\o,\o'}(\pp,\qq)=\hS^{(j,i)}_{\o,\o'}(\qq,\pp)$,
we will only concern the case $i\ge j$.
A bound for $\hS^{(i,j)}_{\o,\o'}$ can be obtained by  explicit computation,
using that 
$$f_i(\kk)\big(1-\c^{-1}_N(\kk)\big)=-\d_{i,N}\big(1-f_N(\kk)\big)
\defi-\d_{i,N}u_N(\kk)\;.
$$
\bd
\item{1.} For $i=j=N$, 
\bea
\hU^{(N,N)}_\o(\qq+\pp,\qq)=
\lft[u_N(\qq){f_N(\qq+\pp)\over D_\o(\qq+\pp)}-
u_N(\qq+\pp){f_N(\qq)\over D_\o(\qq)}\rgt]\bar\c_N(\pp)
\cr\cr
= {u_N(\qq)f_N(\qq+\pp)\over D_\o(\qq)D_\o(\qq+\pp)}\bar\c_N(\pp)D_\o(\pp)
+{f_N(\qq)\over D_{\o}(\qq)} 
\big[u_N(\qq)-u_N(\qq+\pp)\big]\bar\c_N(\pp)
\cr\cr
+{u_N(\qq)\over D_{\o}(\qq)}
\big[f_N(\qq+\pp)-f_N(\qq)\big]\bar\c_N(\pp)
\eea
Therefore we obtain:
\bea
\hS^{(N,N)}_{\o,\o'}(\qq+\pp,\qq)
\defi-\d_{\o,\o'} 
\bar\c_N(\pp){u_N(\qq)f_N(\qq+\pp)\over D_\o(\qq)D_\o(\qq+\pp)}
\cr\cr
+\bar\c_N(\pp){f_N(\qq)\over D_{\o}(\qq)}\int_0^1\!d\t\; 
(\dpr_{\o'}u_N)\big(\qq+\t\pp\big)
\cr\cr
+\bar\c_N(\pp){u_N(\qq)\over D_{\o}(\qq)}\int_0^1\!d\t\; 
(\dpr_{\o'}f_N)\big(\qq+\t\pp\big)
\eea
\item{2.} For $j<N$, using also that $u_N(\qq)f_j(\qq)\=0$
(the support of the two function is disjoint) we have
\bea
\hU^{(N,j)}_\o(\qq+\pp,\qq)=-
\bar\c_N(\pp)u_N(\qq+\pp){f_j(\qq)\over D_\o(\qq)}
\cr\cr
=-
\bar\c_N(\pp)\big[u_N(\qq+\pp)-u_N(\qq)\big]{f_j(\qq)\over D_\o(\qq)}\;.
\eea
Hence:
\be
\hS^{(N,j)}_{\o,\o'}(\qq+\pp,\qq)=\bar\c_N(\pp)
{f_j(\qq)\over D_{\o}(\qq)}\int_0^1\!d\t\; 
(\dpr_{\o'}u_N)\big(\qq+\t\pp\big)
\ee

\item{3.} For $i,j<N$, we have $\hU^{(i,j)}_\o\=0$.
\ed
By inspection, since $|\dpr_\o f_j|\le C\g^{-j}$ 
as well as $|\dpr_\o u_N|\le C\g^{-N}$, we obtain
that  $\dpr^m_{\pp}\dpr^n_{\qq}\hS^{(N,j)}_{\o,\o'}(\pp,\qq)$
is not identically zero if one between $i=N$;
$\qq:f_j(\qq)\not=0$ and $\pp:|\pp|\le 2\g^N$;  in this case we obtain
\bea
\lft|\dpr^m_{\pp}\dpr^n_{\qq}\hS^{(N,j)}_{\o,\o'}(\pp,\qq)\rgt|
\le C_{m,n}\g^{-N(1+m)-j(1+n)}
\eea
The above bounds allow to obtain 
\bea
\int\!{d\pp\;d\qq\over (2\p)^4}\;
\lft|\dpr^m_{\pp}\dpr^n_{\qq}\hS^{(N,j)}_{\o,\o'}(\pp,\qq)\rgt|
\le C'_{m,n}\g^{N(1-m)+j(1-n)}
\eea
from which the former of \pref{61} follows. The analysis for $P^{(N,j)}_{\bar\o,\o}$
is similar:
\bea
\hQ^{(N,j)}_{\o}(\qq+\pp,\qq) 
= \bar\c_N(\pp) u_N(\pp+\qq)\hac_j(\qq)
\cr\cr=
\bar\c_N(\pp) \lft[u_N(\pp+\qq)-u_N(\qq)\rgt] \hac_j(\qq)
\eea 
from which the latter of \pref{61} follows.
\section{Solution of the closed equation}

By \pref{wi} into \pref{sde}, in the limit 
$N\to\io$
%
%
%
%
%
\be\lb{clf}
\dpr_\o S_{\o,\s}(\xx)=
\d(\xx)+\l A_-(\xx) S_{\o,\s}(\xx)
\ee
whose solution is
%
%
\be\lb{sp}
S_{\o,\s}(\xx)=\exp
\lft\{\l\int\!d\zz\; 
\Big[g_\o(\xx-\zz)-g_\o(\zz)\Big]A_-(\zz)
\rgt\} g_{\o}(\xx)
\ee
By \pref{Ae}, we first consider  
\bea\lb{176}
\int\!{d\pp\over (2\p)^2}\; 
e^{-i\xx\cdot\pp}\hg_{\o}(\pp)\hv(\pp)a(\pp)
\cr\cr
=\int\!{d\pp\over (2\p)^2}\;F(\pp)
{e^{-i\xx\cdot\pp}\over (p_0+i\o s(\pp) p_1)(p_0+i\o p_1)}
\eea
where
$$
s(\pp)={1+\n\hv(\pp)\over 1-\n\hv(\pp)}\;,
\qquad
F(\pp)={\hv(\pp)\over \n\hv(\pp)-1}\;.
$$
Indeed \pref{176} is well defined for 
$\xx=0$: we can rewrite it separating the two domains $|\pp|\le 1$ and $|\pp|> 1$.
%
%
The integral on the latter is absolutely convergent, since the decay
of $F(\pp)$ is faster than any power. The integral of the former 
can be written as 
\be\lb{f1}
F(0)\int_{|\pp|\le 1}{d\pp\over (2\p)^2}\;
{1\over (p_0+i\o s  p_1)(p_0+i\o p_1)}+R
\ee
where $R$ is again an absolutely convergent integral; the first
integral can be written
as 
\bea
\int_{|\pp|\le 1}{d\pp\over (2\p)^2}
{1\over (p_0+i\o s p_1)(p_0+i\o p_1)}=
\cr\cr=-
\int_{|\pp|\le 1}{d\pp\over (2\p)^2}
{1\over (i \o p_1+ s p_0)(i \o p_1+p_0)}\cr\cr
=-\int_{p_0^2+s^2 p_1^2\le 1} {d\pp\over (2\p)^2} 
{1\over (p_0+i \o p_1)(p_0+i \o s p_1)}
\eea
hence, the  above integral also equals 
\bea\lb{f111}
\int {d\pp\over (2\p)^2}
{\chi(p_0^2+p_1^2\le 1)-\chi(p_0^2+s^2 p_1^2\le 1)\over
(p_0+i\o s  p_1)(p_0+i\o p_1)}
\eea
which is absolutely convergent since the support of $\chi(p_0^2+p_1^2\le
1)-\chi(p_0^2+s^2 p_1^2\le s^2)$ does not contain 
a neighbourhood of the origin.

Now we discuss \pref{176} for  $\xx\not =0$. It  can be written as
$H_0+ H_1+H_2$, for 
\bea
H_0=F(0)\int\!{d\pp\over (2\p)^2}\;
{e^{-i\xx\cdot\pp}\over (p_0+i\o s p_1)(p_0+i\o p_1)}
\cr\cr
H_1=\int\!{d\pp\over (2\p)^2}\;
{[F(\pp)-F(0)]e^{-i\xx\cdot\pp}\over (p_0+i\o s p_1)(p_0+i\o p_1)}
\cr\cr
H_2=\int\!{d\pp\over (2\p)^2}\;F(\pp)
{e^{-i\xx\cdot\pp}\over (p_0+i\o p_1)}
\lft[{1\over 
(p_0+i\o p_1)}-{1\over 
(p_0+i\o s(\pp) p_1)}\rgt]
\eea
By straightforward computation, 
$H_0$ is given by
\bea
{1\over 2\p(1-\n)(s-1)}
\int_{0}^{+\io}\!{d q_1\over q_1}
\lft[e^{-\big[|x_0|c+ix_1\o{\rm sgn}(x_0)\big]q_1}\rgt.
\cr\cr
-\lft.
e^{-\big[|x_0|+ix_1\o{\rm sgn}(x_0)\big]q_1}\rgt]
\cr\cr
={1\over 4\p\n}
\ln{ x_0+i\o x_1\over x_0s+i\o x_1}\;,
\eea
while both $H_1$ and $H_2$ are vanishing as $\xx\to\io$. Indeed $H_1$ can be written as
\bea
\int_{|\pp|\le N}\!{d\pp\over (2\p)^2}\;[F(\pp)-F(0)]
{e^{-i\xx\cdot\pp}\over (p_0+i\o s p_1)(p_0+i\o p_1)}
\cr\cr+
\int_{|\pp|\ge N}\!{d\pp\over (2\p)^2}\;F(\pp)
{e^{-i\xx\cdot\pp}\over (p_0+i\o s p_1)(p_0+i\o p_1)}
\cr\cr+
F(0)\int_{|\pp|\ge N}\!{d\pp\over (2\p)^2}\;
{e^{-i\xx\cdot\pp}\over (p_0+i\o s p_1)(p_0+i\o p_1)}
\eea
The second and third term are convergent integral, and each of them can be chosen small than
${\e\over 3}$ for $N$ large enough; the first integral is vanishing as $\xx\to \io$.
A similar argument holds for $H_3$.

\end{document}